\documentclass[lettersize,journal]{IEEEtran}
\usepackage{amsmath,amsfonts}
\usepackage{algorithmic}
\usepackage{algorithm}
\usepackage{array}
\usepackage{textcomp}
\usepackage{stfloats}
\usepackage{url}
\usepackage{verbatim}
\usepackage{graphicx}
\usepackage{cite}
\usepackage{multirow}
\usepackage{multicol}
\usepackage{setspace}
\usepackage{gensymb}
\usepackage{makecell}
\usepackage{amssymb}
\usepackage{subfigure}
\usepackage{color}

\begin{document}

\title{SAR Target Image Generation Method\\ Using Azimuth-Controllable Generative Adversarial Network}

\author{Chenwei Wang,~\IEEEmembership{Student Member,~IEEE,}
        Jifang Pei,~\IEEEmembership{Member,~IEEE,}
        Xiaoyu Liu,~\IEEEmembership{Student Member,~IEEE,}
        Yulin Huang,~\IEEEmembership{Senior Member,~IEEE,}
        Deqing Mao,~\IEEEmembership{Student Member,~IEEE,}
        Yin Zhang,~\IEEEmembership{Member,~IEEE,}
        and Jianyu Yang,~\IEEEmembership{Senior Member,~IEEE}
\thanks{This work was supported by the National Natural Science Foundation of China under Grants 61901091 and 61901090.}
\thanks{\emph{Corresponding author: Jifang Pei.}}
\thanks{The authors are with the Department of Electrical Engineering, University of Electronic Science and Technology of China, Chengdu 611731, China (e-mail: peijfstudy@126.com; dbw181101@163.com).}}

\markboth{Journal of \LaTeX\ Class Files,~Vol.~14, No.~8, August~2021}%
{Shell \MakeLowercase{\textit{et al.}}: A Sample Article Using IEEEtran.cls for IEEE Journals}

\maketitle

\begin{abstract}
Sufficient synthetic aperture radar (SAR) target images are very important for the development of researches. However, available SAR target images are often limited in practice, which hinders the progress of SAR application. In this paper, we propose an azimuth-controllable generative adversarial network to generate precise SAR target images with an intermediate azimuth between two given SAR images’ azimuths. This network mainly contains three parts: generator, discriminator, and predictor. Through the proposed specific network structure, the generator can extract and fuse the optimal target features from two input SAR target images to generate SAR target image. Then a similarity discriminator and an azimuth predictor are designed. The similarity discriminator can differentiate the generated SAR target images from the real SAR images to ensure the accuracy of the generated, while the azimuth predictor measures the difference of azimuth between the generated and the desired to ensure the azimuth controllability of the generated. Therefore, the proposed network can generate precise SAR images, and their azimuths can be controlled well by the inputs of the deep network, which can generate the target images in different azimuths to solve the small sample problem to some degree and benefit the researches of SAR images. Extensive experimental results show the superiority of the proposed method in azimuth controllability and accuracy of SAR target image generation.
\end{abstract}

\begin{IEEEkeywords}
Synthetic Aperture Radar (SAR), target image generation, azimuth-controllable, Generative Adversarial Network (GAN), Automatic Target Recognition (ATR), deep learning.
\end{IEEEkeywords}

\section{Introduction}
\IEEEPARstart{S}{ynthetic} aperture radar (SAR) is an important microwave remote sensing system in both mechanism and application, which has the ability to obtain high-resolution images with the pulse compression technology and synthetic aperture principle \cite{curlander1991synthetic,dudgeon1993overview}. It can obtain more distinct information of the target and the scene, for instance, geometry, material, and structure, than optical sensors, infrared sensors, and so on. With the high-resolution coherent imaging capability of all-weather, all day, and penetration, SAR has been widely used in the field of geography, remote sensing, military fields, and so on \cite{wang2022sar,wang2019parking,wang2021deep}.

For the development of the theory and technology of SAR, researches have been carried out in many fields of SAR, such as SAR image despeckling \cite{9323909}, super-resolution \cite{8634345}, target detection, classification, recognition, and multi-sensor image fusion \cite{moreira2013tutorial,addref1}. All these researches are driven by SAR data among which the SAR target images are the most important. For example, in the field of SAR image despeckling \cite{8899122}, SAR images are necessary for the researches on the SAR speckle characteristics and the despeckling algorithm of both traditional methods and deep learning \cite{dai2004bayesian}. In the field of SAR multi-sensor image fusion \cite{9323085,addref3,addref4} and pixel image fusion for SAR and optical images \cite{addref2} all require high-quality SAR images for better fusing the scene information and the interpretation of SAR scene images \cite{simone2002image}. As the most representative, in the field of SAR automatic target recognition (ATR) \cite{wang2023entropy,wang2023sar,wang2022semi,wang2020deep,wang2021multiview,wang2020multi}, a great quantity of SAR target images is necessary for the acquirement of target features, improvement of the recognition ratio, and promotion of the practical application of SAR ATR \cite{ross1998standard}.

However, in the actual situation, an abundant number of SAR images is lacking, and SAR image acquisition is difficult and consumes resources. Even if there are some SAR images, they are likely obtained by different imaging conditions, such as the band, platform, azimuth, and so on, and these SAR images can not contain enough information of the scene or target for the researches of SAR fields. The insufficiency of SAR image data or lack of the characteristics of scene or target in SAR images have become a great obstacle of almost all SAR fields and hinders the progress of SAR application \cite{balz2009hybrid}.

To solve this problem, many kinds of research are carried out in recent years \cite{EliMRec,SLMRec,BundleGT,DBLP:conf/emnlp/ZhaoWLSZ022,DBLP:conf/acl/Zhao00WZZC23,wang2022recognition,wang2022global}. And there are mainly three types of SAR target image acquirement: measured data collection, electromagnetic simulation, and sample augmentation \cite{franceschetti1992saras,zhang2018multiple}. First of all, measured data collection can obtain the SAR target images under different actual scenarios with different platforms. These acquired data are the most authentic and effective. However, this acquirement will consume massive resources of human, material and time, and the number of the acquired SAR target images in each experiment is often limited. The result is that it cannot be used as a cost-effective way for application to obtain enough SAR data. 

Through the 3-D modelling of the target and electromagnetic calculation imaging, the electromagnetic simulation is, although not as accurate as the real SAR data, comparatively accurate \cite{collins2009modeling}. The results of the simulation SAR target images are related to how precise the 3-D models and electromagnetic calculation methods are. But the more accurate the 3-D model and electromagnetic calculation method are, the greater the computation will be and the slower the compute process is, which can lead to the massive consumption of the time resource. Besides, when every single different radar parameter gets changed, the computation of electromagnetic simulation needs to start from scratch without using prior knowledge of existing simulation data. 

As a result, although electromagnetic simulation is an alternative approach, it cannot also be an effective way to solve the lack of radar data amount. Last of all, sample augmentation, mainly employed in the field of SAR ATR, is to increase the diversity of the SAR sample and avoid overfitting of the classifier, such as translation, rotation, adding-noise, and so on \cite{zhang2018data}. Luo et al. proposed a synthetic minority class data method for improving imbalanced SAR target recognition using the generative adversarial network (GAN) \cite{9323439}. However, these methods are only the augmentation from the view of image processing, and many augmented images do not conform to the law of radar imaging and do not contain new information. Therefore, it can not increase the intrinsic information of the target essentially.

In recent years, when deep learning has been applied in signal and image processing fields and demonstrated its superior performance, lots of excellent scholars mainly focus on the SAR image generation and have proposed several deep learning methods with outstanding results \cite{el2016automatic,wilmanski2016modern,gong2017feature,ganadd3}. For example, Guo et al. \cite{guo2017synthetic} proposed a conditional generative adversarial net (CGAN) with a clutter normalization method to ease the model collapse during the generation of SAR images. Cui et al. \cite{cui2019image} proposed a deep convolutional GAN (DCGAN) to generate SAR images with random azimuths and employed an azimuth discriminator to filter the desire generated images with the azimuths close to specific angles. Jiang et al. \cite{8518792} proposed a Gabor-Deep Convolutional Neural Networks (G-DCNNs) which is a method of data augmentation with Gabor filter in DCNNs for SAR ATR. It overcame the severe overfitting due to limited SAR image training data when applying DCNNs. Zheng et al. \cite{zheng2019semi} proposed a multi-discriminator GAN with a label smoothing regularization to generate SAR target images with unclear types. Du et al. \cite{9387399} proposed a multiconstraint GAN (MCGAN) to generate high-quality multicategory SAR images to address the poor image quality problem. Mao et al. \cite{ganadd2} combined Constrained Naive Generative Adversarial Networks (CN-GAN) with least squares generative adversarial networks and image-to-image translation to address the problem of low signal-to-clutter-noise ratio, model instability and the excessive freedom degree of the output, which appeared in conventional GAN. Saha et al. \cite{ganadd1} employed transfer learning framework using a cycle-consistent generative adversarial network (CycleGAN) to train for the suboptimal task of transcoding SAR images into optical images. These existing deep learning SAR image generators greatly promote the research of SAR image acquirement.

However, most of the current SAR image generation methods are just the augmentation of the SAR dataset and generated from random noise, which means these methods can only generate abundant SAR images without controlling the azimuths. When the azimuth distribution of the SAR dataset is not balanced, the generated SAR image dataset by current generation methods are concentrated around certain azimuths. In practice, the features of the target in the SAR image are changing when the azimuth of the SAR image are different. And the lacking of azimuth in the SAR image dataset is actually  equivalent to the lacking in the features of target, which can negatively influence the recognition results and other researches \cite{pei2017sar}. Therefore, the azimuth-controllable generation of the SAR target images is beneficial and necessary for the improvement of target recognition and other researches. It is still a gap between the practical demands and the current methods.

Therefore, we proposed an azimuth-controllable generative adversarial network in this paper, which can generate precise SAR images with an intermediate azimuth between two given SAR images’ azimuths. The azimuth-controllable SAR target image generation method mainly consists of two parts: the generator and the discriminator. First, through the proposed specific topological structure of the dual-input parallel SAR image input block, the generator can extract the target feature from the two input SAR images with different azimuths. Second, the specific topological structure of the similarity discriminator and the azimuth predictor is designed. The similarity discriminator can distinguish the generated SAR images from the real SAR images, and the azimuth predictor can obtain the distance between the generated image’s azimuth and the desired image’s azimuth. Then, the proposed azimuth-controllable SAR target image generation network can generate precise SAR images from two input SAR images and the azimuths of the generated SAR images can be controlled by the azimuths of the two input SAR images. These generated SAR images of different azimuths can help to solve the small sample problem in some degree and benefit the research of target characteristic in SAR images.
The main contributions compared with available works are the following:

(1) We proposed a framework of azimuth-controllable GAN, which construct an adversarial game between the generator and similarity discriminator with the azimuth predictor, then learns the distribution of feature and azimuth from two input SAR images with azimuth, and generate precise SAR images with an intermediate azimuth between the two adjacent azimuths. Besides, the proposed corresponding data formation method make the azimuth-controllable GAN can achieve the general process of gradually learning the low-dimension manifold of SAR image azimuth.

(2) We proposed a generator of a special topological structure and, azimuth predictor and similarity discriminator. This generator can generate precise SAR images with azimuth controllability. And it can solve the model collapse to make the training stable, by the special structure and input SAR images. The azimuth predictor and similarity discriminator provide accurate information of optimization by the structure and module in discriminator and an adversarial mechanism in the aspect of similarity and azimuth separately. Furthermore, to make the training more stable, the loss of azimuth predictor takes the mean square error to replace the logarithmic error, and the loss of the azimuth predictor takes Wasserstein distance to relieve the model collapse.

(3) In the experiment, the generated SAR images not only achieve high similarity qualitatively and quantitatively but also promote greatly the recognition performance in a small dataset.

The remainder of this paper is organized as follows. An overview of the azimuth-controllable SAR target image generation network is presented in Section 2. Section 3 evaluates the performance of our proposed SAR target image generation network with experiments, and Section 4 gives a brief conclusion.

\section{Proposed Azimuth-controllable Generative Adversarial Network}

\begin{figure*}[!htb]
    \centering
    \includegraphics[width=1\textwidth]{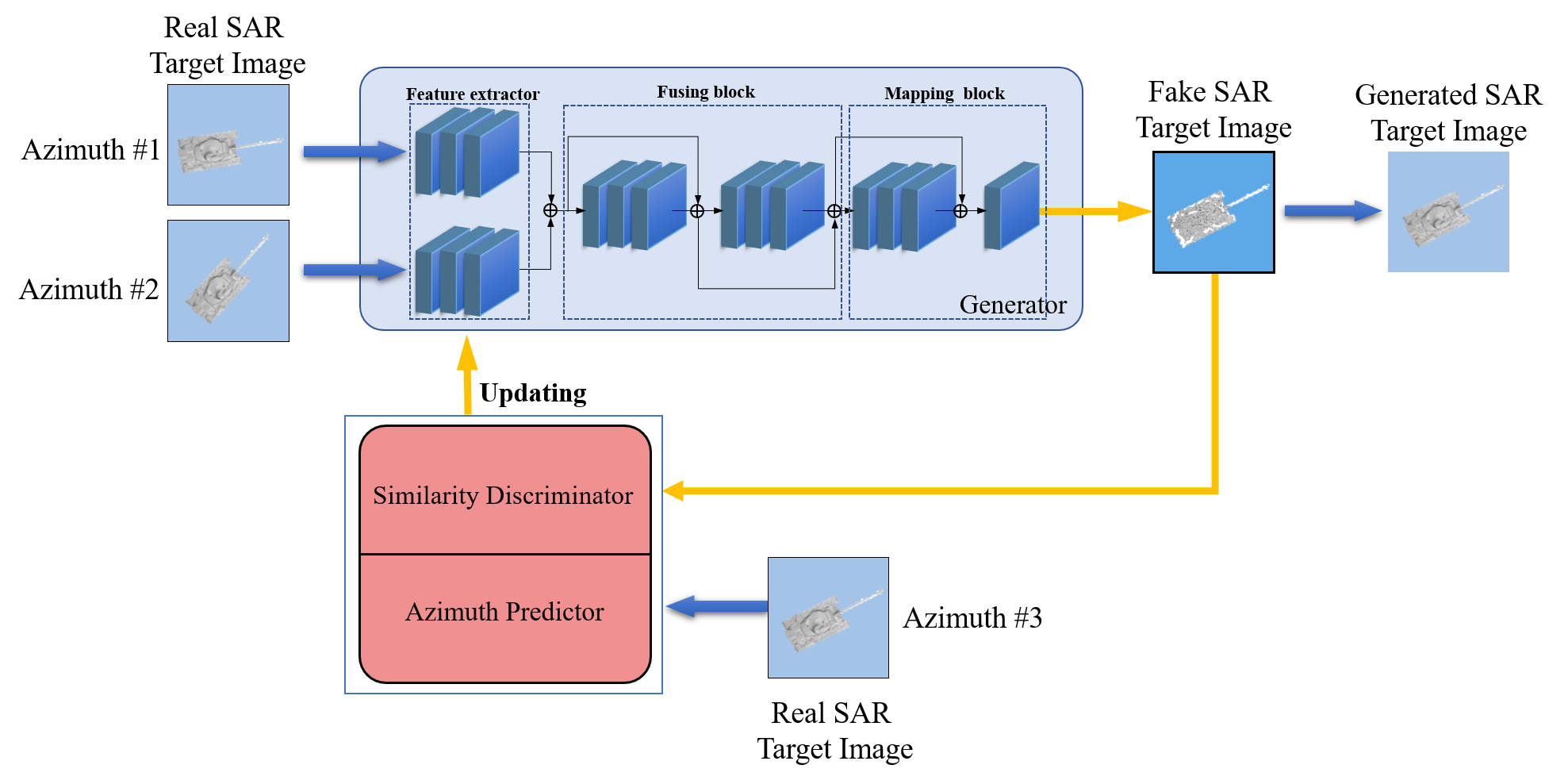}
    \caption{Schematic of azimuth-controllable SAR target image generation network.}\label{figure1}
\end{figure*}

In this section, the framework of the proposed azimuth-controllable SAR target image generation network is presented. First, we elucidate the framework of the azimuth-controllable SAR target image generation network. Then, the specific structures of the generator, discriminator, predictor in the proposed network will be presented.

\subsection{Framework of Proposed Network}

According to the manifold learning theory, data in the high-dimension space is often mapped from a low-dimension manifold and so do the targets in SAR image. The targets in SAR image in continuous azimuth can be mapped onto a low-dimension manifold \cite{pei2016sar}, which means the distribution of target in SAR image are stable and learnable. Therefore, we proposed the azimuth-controllable SAR target image generation network, through designing the specific architecture and input form, to gradually learn the low-dimension manifold of the target in continuous azimuth and generate precise SAR target images in specific azimuth. The general process of gradually learning the low-dimension manifold by the proposed azimuth-controllable SAR target image generation network can be described as follows.

Given two SAR target images, ${{\bf{I}}_{\rm{1}}}$ and ${{\bf{I}}_{\rm{2}}}$, with adjacent azimuths ${\theta _{\rm{1}}}$ and ${\theta _{\rm{2}}}$, as the input of the generator $G$, the fake SAR target image ${{\bf{I}}_g}$ with the azimuth ${\theta _1} < {\theta _g} < {\theta _2}$ will be generated by $G$. However, in this initial state, the fake SAR target image ${{{\bf{I}}}_{g}}$ has little same information as the real SAR target image ${{{\bf{I}}}_{r}}$ with ${{\theta }_{g}}$. Then, the two images, fake ${{{\bf{I}}}_{g}}$ and real ${{{\bf{I}}}_{r}}$, are inputs to the discriminator and predictor, ${{D}_{o}}$ and ${{D}_{a}}$. The similarity discriminator ${{D}_{o}}$ will determine whether each image is real or fake, and the azimuth predictor ${{D}_{a}}$ will predict the azimuths of the two images. Through adversarial training, the generator and discriminator are both optimized \cite{goodfellow2014generative}. Finally, the SAR target image ${{{\bf{I}}}_{g}}$ with the azimuth ${{\theta }_{1}}<{{\theta }_{g}}<{{\theta }_{2}}$ will be generated with high quality. After the rough process of describing how the proposed azimuth-controllable SAR target image generation network gradually learns the low-dimension manifold, the novel framework of the generator, discriminator, and predictor will be present to show how each part of the azimuth-controllable SAR target image generation network completes its function.

As shown in Fig.\ref{figure1}, the framework of the proposed azimuth-controllable SAR target image generation network consists of two parts, a generator, and a discriminator. For the generator, a specific topologic architecture with two parallel input block is designed to achieve the goal of extracting the target features from the two different input SAR images with adjacent azimuths, ${{\theta }_{1}}$ and ${{\theta }_{2}}$. Besides, the cascaded residual blocks are also adopted in the architecture to preserve and fuse adaptively the extracted target features along the pipeline \cite{he2016deep}. For the discriminator, similarity discriminator ${{D}_{o}}$ and azimuth predictor ${{D}_{a}}$ are designed. The similarity discriminator is designed to distinguish the generated SAR images from the real SAR images and provide the optimization direction of the image distribution. As for the azimuth predictor ${{D}_{a}}$, it can acquire accurate azimuth of SAR target images and provide the optimization direction of the azimuth, which can lead to the azimuth controllability of the generated SAR image. The output fake SAR target image represents the generated image that is not good enough during the training process, and is considered as a fake image by the discriminator. When the training is basically completed and the features and azimuths of the generated image are fairly accurate, the generated SAR target image is the image generated by the network after sufficient training, which are the azimuth-controllable images we want to get.

Through the input form and architecture of the generator, discriminator, and predictor, the proposed azimuth-controllable SAR target image generation network can acquire a steadier training process than the other architecture of GAN for SAR image generation and generate more precise SAR target images with azimuth controllability.

\subsection{Specific Implementation of Generator, Discriminator and Predictor}

\begin{figure*}[!htb]
    \centering
    \includegraphics[width=1\textwidth]{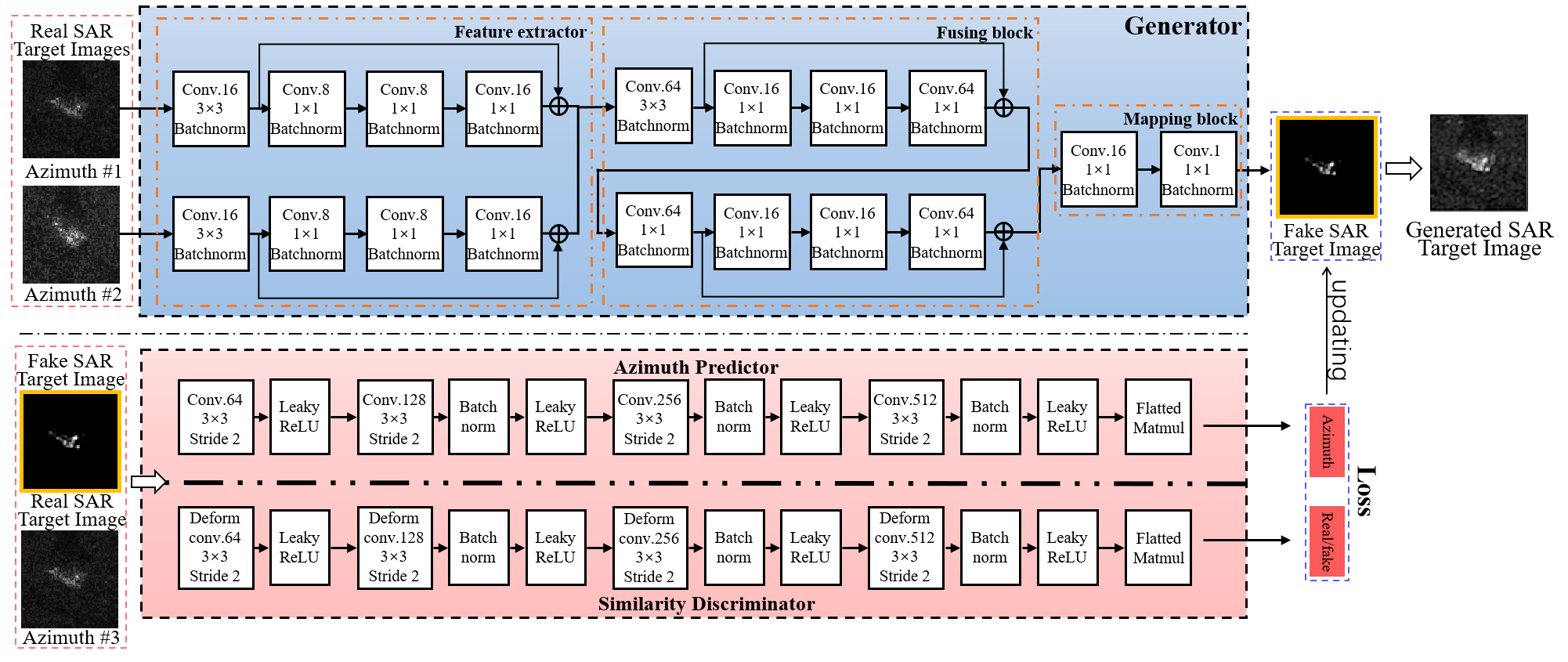}
    \caption{Structures of generator, discriminator, and predictor.}\label{figure2}
\end{figure*}

To generate the SAR target images with the controllable azimuth from the input SAR images with the adjacent azimuths, the structures and losses of the generator, discriminator, and predictor are designed as shown in Fig.\ref{figure2}.

To extract the azimuth-specific information and learn the manifold of the target in continuous azimuth, a special topology structure is proposed with three parts, two parallel input blocks ${{B}_{pi1}}\left( \cdot  \right)$ and ${{B}_{pi2}}\left( \cdot  \right)$, an information-fusing block ${{B}_{if}}\left( \cdot  \right)$ , and a mapping block ${{B}_{m}}\left( \cdot  \right)$. Two SAR images ${{{\bf{I}}}_{1}}$ and ${{{\bf{I}}}_{2}}$ are input to the two parallel input blocks ${{B}_{pi1}}\left( \cdot  \right)$ and ${{B}_{pi2}}\left( \cdot  \right)$ separately and the information of the target feature is learned separately from the two input images with adjacent azimuths by
{
\setlength\abovedisplayskip{5pt}
\setlength\belowdisplayskip{2.5pt}
\begin{flalign}
{{\bf{l}}_{pi1}} = {B_{pi1}}\left( {{{\bf{I}}_1}} \right)
\end{flalign}}
{
\setlength\abovedisplayskip{2.5pt}
\begin{flalign}
{{\bf{l}}_{pi2}} = {B_{pi2}}\left( {{{\bf{I}}_2}} \right)
\end{flalign}}
where ${{\mathbf{l}}_{pi1}}$ and ${{\mathbf{l}}_{pi2}}$ denote the target features under different azimuths of the two inputs. Then the information-fusing block ${{B}_{if}}\left( \cdot  \right)$ is employed to obtain and retain the target information of the SAR target image with the azimuth ${{\theta }_{1}}<{{\theta }_{g}}<{{\theta }_{2}}$ adaptively by mapping the information between ${{\mathbf{l}}_{pi1}}$ and ${{\mathbf{l}}_{pi2}}$, which can be defined as

{
\setlength\abovedisplayskip{1.5pt}
\begin{flalign}
{{\bf{R}}_{{\mathop{\rm int}} erp}} = {B_{if}}\left( {{{\bf{l}}_{pi1}},{{\bf{l}}_{pi2}}} \right)
\end{flalign}}
where ${{\mathbf{R}}_{\operatorname{int}erp}}$ denotes the mapped information of the SAR target image with azimuth ${{\theta }_{1}}<{{\theta }_{g}}<{{\theta }_{2}}$. At last, combining the target information, the mapping block ${{B}_{m}}\left( \cdot  \right)$ can map the information from high dimension into the 2-dimension image space to generate the final precise SAR target images. Through this topology structure, the generator can obtain the capability of mapping the two input SAR images ${{{\bf{I}}}_{1}}$ and ${{{\bf{I}}}_{2}}$ with azimuth ${{\theta }_{1}}$ and ${{\theta }_{2}}$ to the SAR image ${{{\bf{I}}}_{g}}$ with the azimuth ${{\theta }_{1}}<{{\theta }_{g}}<{{\theta }_{2}}$.

To maximize the retention of target information, the residual block is adopted in all three blocks, and batch normalization \cite{ioffe2015batch} is adopted in each layer.

Then, the similarity discriminator and azimuth predictor are designed in the similar structure. They both use the strided convolutions instead of pooling layers to allow the network to learn its own spatial downsampling and leaky rectified linear unit (LReLU) activation \cite{maas2013rectifier} for all layers. Meanwhile, batch normalization is adopted and all dense layers are removed for deeper architectures. For the similarity discriminator, the last convolution layer is flatted and then fed into a vector multiplication. After updating the parameter of the similarity discriminator, the weigh clipping is adopted to ensure the objective function as earth-mover (EM) distance \cite{arjovsky2017wasserstein}. For the azimuth predictor, considering the sensitivity for the azimuth, the deformable convolution \cite{dai2017deformable} is adopted in the azimuth predictor. This deformable convolution has the capability of obtaining the target contour and direction by changing the shapes of the convolution kernels \cite{dai2017deformable}. And the last convolutional layer of the azimuth predictor is flattened without activation as it is just a vector multiplication.

As described above, the training process of the proposed azimuth-controllable SAR target image generation network is a competition among the generator, discriminator, and predictor. More formally, the total value function of the proposed azimuth-controllable SAR target image generation network can be expressed as:

\begin{flalign}
\left\{
\begin{array}{l}
{{F}_{1}}=\underset{G}{\mathop{\min }}\,\underset{{{D}_{0}}}{\mathop{\max }}\,\big( {{E}_{\mathbf{I}\sim{{P}_{data}}\left( \mathbf{I} \right)}}\left[ \log \left( {{D}_{0}}\left( {{\mathbf{I}}_{r}} \right) \right) \right] \\
\text{                                     }+{{E}_{\left( {{\mathbf{I}}_{1}},{{\mathbf{I}}_{2}} \right)\sim{{P}_{data}}\left( \mathbf{I} \right)}}\left[ \log \left( 1-{{D}_{0}}\left( G\left( {{\mathbf{I}}_{1}},{{\mathbf{I}}_{2}} \right) \right) \right) \right] \\
\text{                                     }+Eu\left( {{D}_{a}}\left( G\left( {{\mathbf{I}}_{1}},{{\mathbf{I}}_{2}} \right) \right),{{\theta }_{g}} \right) \big) \\
{F_2} = \mathop {\min }\limits_{{D_a}} Eu\left( {{D_a}\left( {{{\mathbf{I}}_r}} \right),{\theta _g}} \right)
\end{array}
\right.
\end{flalign}
where $E$ denotes the expectation operator, $Eu$ denotes the Euclidean Distance, ${\bf{I}}_r$ denotes the input real SAR images of the discriminator and predictor, ${{\bf{I}}_{1}}$ and ${{\bf{I}}_{2}}$ denotes the two input real SAR images of the generator, and ${{P}_{data}}\left( {\bf{I}} \right)$ denotes the distribution of the real SAR images, $\theta_{g}$ means the azimuth of the input real SAR images of the predictor. The first two terms of ${{F}_{1}}$ are to ensure that the generated SAR images can be accurately distinguished by the discriminator. The third term of ${{F}_{1}}$ is for the generator to generate the images with the expected azimuth. As for the term of ${{F}_{2}}$, it is an azimuth loss for the azimuth predictor, which intends to minimize the distance of the azimuth between the real SAR images and the generated SAR images.

Therefore, the loss of the similarity discriminator can be present by

\begin{flalign}
\begin{split}
{L_{{D_o}}}\left( {G,D_0} \right) = & - ({E_{{\bf{I}}_r \sim {P_{data}}\left( {\bf{I}} \right)}}\left[ {\log \left( {D_0\left( {\bf{I}}_r \right)} \right)} \right]\\
& + {E_{\left( {{{\bf{I}}_1},{{\bf{I}}_2}} \right) \sim {P_{data}}\left( {\bf{I}} \right)}}\left[ {\log \left( {1 - D_0\left( {G\left( {{{\bf{I}}_1},{{\bf{I}}_2}} \right)} \right)} \right)} \right])
\end{split}
\end{flalign}
where $E$ denotes the expectation operator, ${\bf{I}}_r$ denotes the input real SAR images of the discriminator and predictor, ${{\bf{I}}_{1}}$ and ${{\bf{I}}_{2}}$ denotes the two input real SAR images of the generator, and ${{P}_{data}}\left( {\bf{I}} \right)$ denotes the distribution of the real SAR images.

For the azimuth predictor, the mean square error is proposed. The loss of the azimuth predictor can be present by

\begin{flalign}
\begin{split}
{L_{{D_a}}}\left( {{{\bf{I}}_r},{\theta _g}} \right) = &Eu\left( {{D_a}\left( {{{\bf{I}}_r}} \right),{\theta _g}} \right)\\
=  & \sqrt {{{\left( {{D_a}\left( {{{\bf{I}}_r}} \right) - {\theta _g}} \right)}^2}}
\end{split}
\end{flalign}
where ${\bf{I}}_{r}$ denotes the input real SAR images of the discriminator and predictor and $\theta {}_{g}$ means the azimuth of the input real SAR images of the predictor.

As for the loss function of the generator, wasserstein-1 distance is used to replace the Jensen Shannon divergence of traditional GAN. It can be defined by

\begin{flalign}
\begin{split}
{L_G}\left( {G,D_0} \right) = &  {E_{\left( {{{\bf{I}}_1},{{\bf{I}}_2}} \right) \sim {P_{data}}\left( {\bf{I}} \right)}}\left[ {\log \left( {1 - D_0\left( {G\left( {{{\bf{I}}_1},{{\bf{I}}_2}} \right)} \right)} \right)} \right]\\
& + \delta {L_{{D_a}}}\left( {{{\bf{I}}_1},{{\bf{I}}_2},{\theta _g}} \right)\\
= &  {E_{\left( {{{\bf{I}}_1},{{\bf{I}}_2}} \right) \sim {P_{data}}\left( {\bf{I}}\right)}}\left[ {\log \left( {1 - D_0\left( {G\left( {{{\bf{I}}_1},{{\bf{I}}_2}} \right)} \right)} \right)} \right]\\
& + \sqrt {{{\left( {{D_a}\left( {G\left( {{{\bf{I}}_1},{{\bf{I}}_2}} \right)} \right) - {\theta _g}} \right)}^2}}
\end{split}
\end{flalign}

As the proposed azimuth-controllable SAR target image generation network is training, the generator, discriminator, and predictor are updated alternately to be optimized. Therefore, while the discriminator and predictor can recognize the images and predict the azimuth more accurately, the generator can make the generated SAR images close to the real images in similarity and azimuth.

The exact steps of the data formation can be described as follow. For each target type, suppose the SAR images ${{\mathbf{X}}^{\left( l \right)}}=\left\{ {{\mathbf{x}}_{1}},{{\mathbf{x}}_{2}},\ldots ,{{\mathbf{x}}_{k}} \right\}$ of $lth$ target type is sort as the azimuth ${{\mathbf{\theta }}^{\left( l \right)}}=\left\{ {{\theta }_{1}},{{\theta }_{2}},\ldots ,{{\theta }_{k}} \right\}$ increases. For an image ${{\mathbf{x}}_{k}}$ in ${{\mathbf{X}}^{\left( l \right)}}=\left\{ {{\mathbf{x}}_{k}},{{\mathbf{x}}_{k+1}},\ldots ,{{\mathbf{x}}_{n}} \right\}$ as the half of input of the generator, with a set azimuth interval $\delta $, the other half of input is set by finding a images ${{\mathbf{x}}_{kf}}$ with the closest azimuth to ${{\theta }_{k}}+\delta $ in the sub-dataset $\left\{ {{\mathbf{x}}_{k+1}},\ldots ,{{\mathbf{x}}_{n}} \right\}$. Then the input of the generator is set as $\left\{ {{\mathbf{x}}_{k}},{{\mathbf{x}}_{kf}} \right\}$ and the input of the discriminator is set by finding some images ${{\mathbf{x}}_{kd}}$ with the azimuths $\frac{{{\theta }_{k}}+{{\theta }_{kf}}}{2}\text{-}\varepsilon \le {{\theta }_{\text{kd}}}\le \frac{{{\theta }_{k}}+{{\theta }_{kf}}}{2}\text{+}\varepsilon $ in the sub-dataset $\left\{ {{\mathbf{x}}_{k1}},\ldots ,{{\mathbf{x}}_{kf-1}},{{\mathbf{x}}_{kf+1}}\ldots ,{{\mathbf{x}}_{n}} \right\}$. At the time, one combination of the inputs to the generator and discriminator are set as $\left\{ {{\mathbf{x}}_{k}},{{\mathbf{x}}_{kf}} \right\}$ and $\left\{ {{\mathbf{x}}_{kd}} \right\}$. Then next combination starts in the subset without $\left\{ {{\mathbf{x}}_{k}},{{\mathbf{x}}_{kd}},{{\mathbf{x}}_{kf}} \right\}$.

Through the network design above and the training process, the parameters of the network are optimal. The proposed azimuth-controllable SAR target image generation network can generate precise SAR images and control the azimuth of the generated SAR images.

\section{Experiments and Results}

\begin{figure*}[!htb]
    \centering
        \includegraphics[width=1\textwidth]{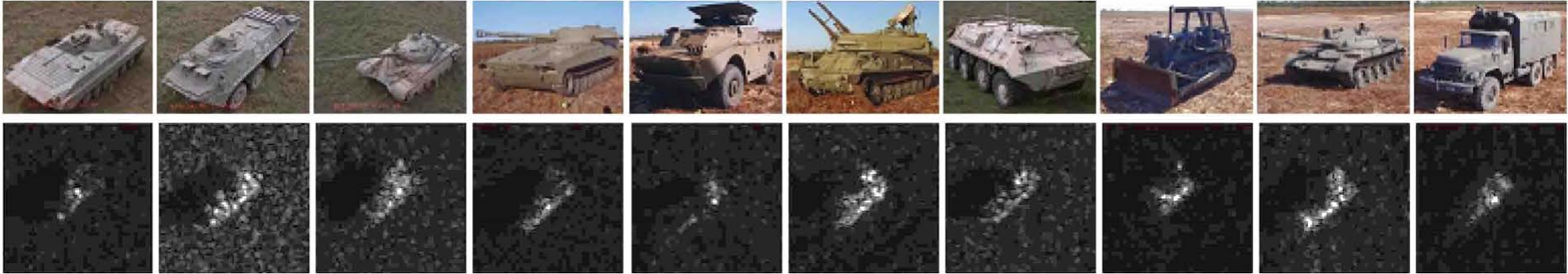}
    \caption{SAR images and corresponding optical images of targets at similar aspect angles.}\label{figure3}
\end{figure*}

In this section, the performance of the proposed azimuth-controllable SAR target image generation network will be evaluated. The Moving and Stationary Target Acquisition and Recognition (MSTAR) is used to evaluate the whole network, and the image data of the training and testing will be firstly introduced in detail. Then, the performance of the azimuth-controllable SAR target image generation network for generating the stable and precise SAR target images will be evaluated with different azimuth intervals, the quality of the generated SAR target images will be presented as well. Finally, as the evaluation for an application of the generated SAR target images, the improvement and comparison of SAR ATR under both standard operating conditions (SOC) and extended operating conditions (EOC) will be presented \cite{ross1998standard}.

\subsection{Dataset and Configuration}

The experiment dataset used to evaluate our proposed azimuth-controllable SAR target image generation network is collected from the MSTAR program. This dataset is released by the U. S. Defense Advanced Research Projects Agency and the Air Force Research Laboratory. The dataset is collected using the Sandia National Laboratory STARLOS sensor platform. As a benchmark dataset for SAR ATR performance assessment, this dataset has a significant quantity of SAR images containing ten different classes of ground targets (tank: T62, T72, rocket launcher: 2S1, truck: ZIL131, armored personnel carrier: BTR70, BTR60, BRDM2, BMP2, air defense unit: ZSU23/4 and bulldozer: D7), which are captured as 1-ft resolution X-band SAR images with full azimuth coverage (in the range of 0\textdegree to 360\textdegree). These SAR images are collected under varying operating conditions, such as different aspect angles, depression angles, and serial numbers. The SAR images and corresponding optical images of the target at similar aspect angles are depicted in Fig.\ref{figure3} \cite{ross1998standard}.

On the basis of the proposed azimuth-controllable GAN, a specific implement is employed to evaluate the proposed framework, which  is present in Fig.2. The size of the input SAR images is $88\times88$, the stride size of every convolutional layer is $1\times1$ in the generator. Other hyper parameters in our network instances are shown in Fig 2. During the training, the range of weight clipping are $-0.01$ to $0.01$ for the discriminator. To balance the adversarial game among the generator, discriminator, and predictor, the discriminator and predictor are optimized 25 times when the generator is optimized once.

The proposed methods is tested and evaluated on a computer with Inter Core I7-9700K at 3.6GHz CPU, Gefore GTX 1080ti GPU with two 16GB memories. The proposed method is implemented using the open-source TensorFlow framework.

\subsection{Evaluation of Generation Ability of Proposed Network}

In this subsection, after the introduction of dataset configuration for the evaluation of the generation ability, the generation ability of the proposed azimuth-controllable SAR target image generation network will be evaluated from the qualitative and quantitative respects. To show the visual similarity, the selected generated images of the ten targets will be present with the real images with close azimuth. For the quantitative similarity, the metrics of image similarity will be employed.

\subsubsection{Dataset Configuration}

The SAR image at 17\textdegree depression angle is set as the training and testing dataset for the generation ability. And the original number of the SAR images in the MSTAR dataset is listed in Table \ref{table1}. Besides, the original dataset is divided into the training and testing dataset as 1:1, whose numbers are also listed in Table \ref{table1} together.

\begin{table*}[!t]
\renewcommand{\arraystretch}{1.3}
\caption{Original number of MSTAR and training and testing dataset}\label{table1}
\begin{tabular}{p{3.3cm}<{\centering}|p{3cm}<{\centering}|p{3cm}<{\centering}|p{3cm}<{\centering}|p{3cm}<{\centering}}
\hline \hline
  & \multicolumn{2}{c|}{Original} & Training & Testing \\
\hline
Class  & Depression & Number & Number & Number  \\
\hline
BMP2-9563  & 17\textdegree & 233 & 117 & 116  \\
\hline
BTR70-c71 & 17\textdegree & 233 & 117&	116  \\
\hline
T72-132 & 17\textdegree & 232 & 116	&116  \\
\hline
BTR60-7532 & 17\textdegree & 256 & 128	&128  \\
\hline
2S1-b01 & 17\textdegree & 299 & 150	&149  \\
\hline
BRDM2-E71 & 17\textdegree & 298 & 149	&149  \\
\hline
D7-92 & 17\textdegree & 299 & 150	&149  \\
\hline
T62-A51 & 17\textdegree & 299 & 150	&149  \\
\hline
ZIL131-E12& 17\textdegree & 299 & 150	&149  \\
\hline
ZSU234-d08 & 17\textdegree & 299 & 150	&149  \\
\hline \hline
\end{tabular}
\end{table*}

\begin{table}[!htb]
\small
\centering
\begin{spacing}{1.19}
\caption{Number of training and testing dataset for the evaluation of the generation ability}\label{table2}
\begin{tabular}{p{2.4cm}<{\centering}p{2.4cm}<{\centering}p{2.4cm}<{\centering}}
\hline \hline
\multirow{2}{*}{Azimuth interval} & Training  & Testing \\
& Combination  & Combination \\
\hline
5\textdegree&	550	&549  \\
\hline
10\textdegree&308&	307 \\
\hline
15\textdegree&	214	&214  \\
\hline
20\textdegree&	167 &167 \\
\hline \hline
\end{tabular}
\end{spacing}
\end{table}

As described above in Section II. B, the input form is combined separately as the azimuth interval are 5\textdegree, 10\textdegree, 15\textdegree, and 20\textdegree. After setting the azimuth interval as a certain value, and dividing the MSTAR dataset into the training and testing dataset, the training and testing dataset go through the data selection described above to get the training and testing combination which is listed in Table.2. For enough training data of the proposed GAN, the data listed in Table \ref{table2} is augmented 10 times by randomly sampling ten $88\times88$ SAR image chips from one original  $128\times128$ SAR image, which ensures the target complete []. Finally, the training and testing data are used to evaluate the performance of the azimuth-controllable SAR target image generation network.

\subsubsection{Evaluation in Visual Similarity}

\begin{figure*}[!htb]
\begin{center}
    \subfigure[]{\label{1.1}\includegraphics[width=5.8in]{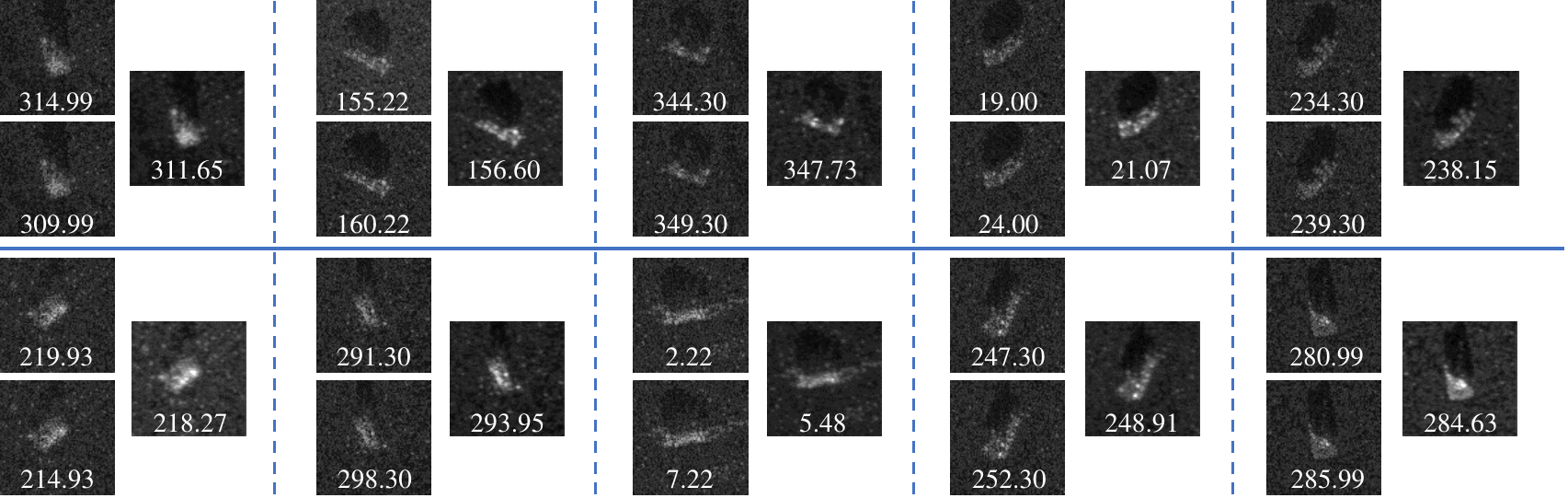}}
    \subfigure[]{\label{1.2}\includegraphics[width=5.8in]{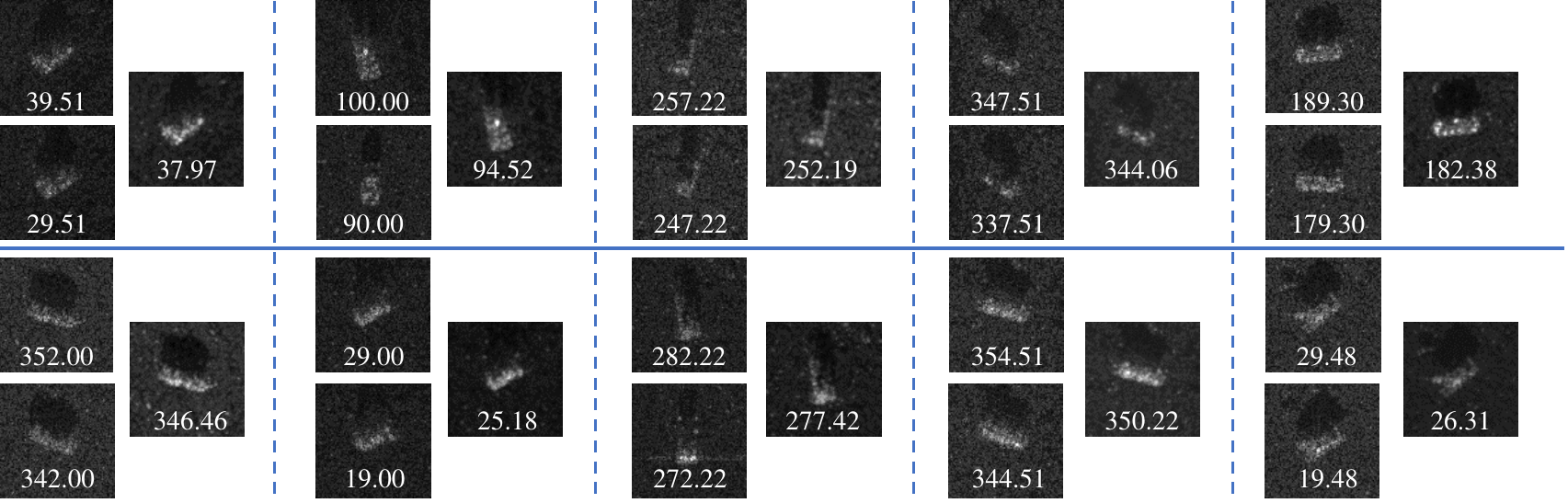}}
    \subfigure[]{\label{1.3}\includegraphics[width=5.8in]{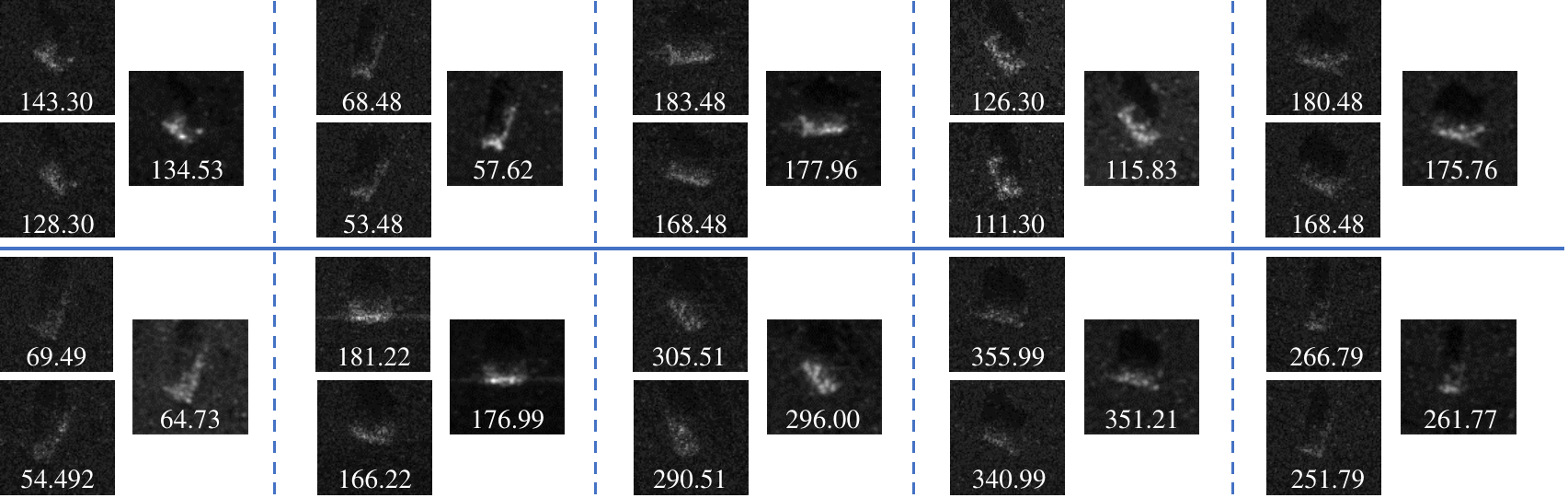}}
    \subfigure[]{\label{1.4}\includegraphics[width=5.8in]{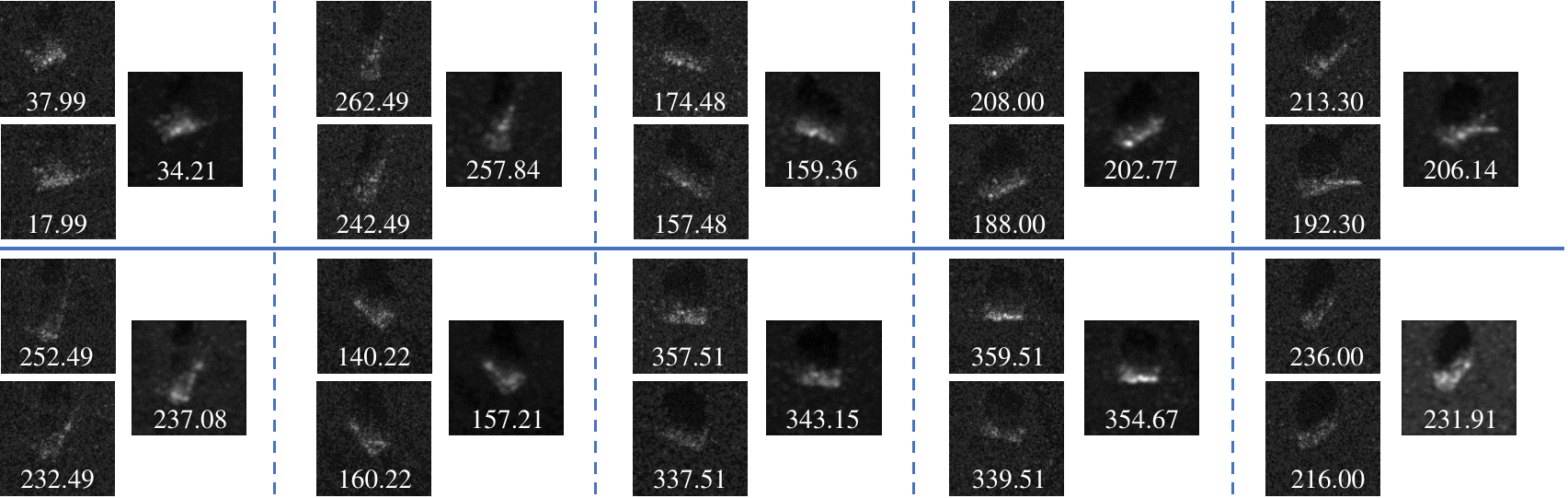}}\\
\end{center}
\caption{10 randomly chosen generated SAR images and the input real SAR images of the generator. In ten combinations of three images in each subfigure, the left two images are the input real SAR images of the generator and the right one is the generated. (a) generated SAR images and the input real SAR images with 5\textdegree azimuth interval, (b) with 10\textdegree azimuth interval, (c) with 15\textdegree azimuth interval, and (d) with 20\textdegree azimuth interval.}
\label{figure4}
\end{figure*}

To evaluate the generation capability of the proposed azimuth-controllable SAR target image generation network, the generated SAR images of 10 targets are randomly chosen in azimuth ranging from 0\textdegree to 360\textdegree, and the input real SAR images of the generator are presented in Fig.\ref{figure4}. Besides, with the decreasing number of training and testing datasets, the generated SAR images will be presented together to evaluate the robustness under the limited training dataset.

As shown in Fig.\ref{figure4}, by comparing the generated SAR images with the two input of the generator, it can be seen that the azimuths of the generated SAR images are intermediate between the azimuths of the input SAR images. When the target type is varying, the azimuths of the generated SAR images are in the desired range. Besides, when the azimuth interval is 5\textdegree, 10\textdegree, 15\textdegree and 20\textdegree respectively, the azimuths of the generated SAR images can still stay in the desired azimuth range. In conclusion, it is clear that the proposed network has azimuth controllability of the generated SAR images.

\begin{figure*}[!htb]
\begin{center}
\subfigure[]{\label{1.1}\includegraphics[width=6in]{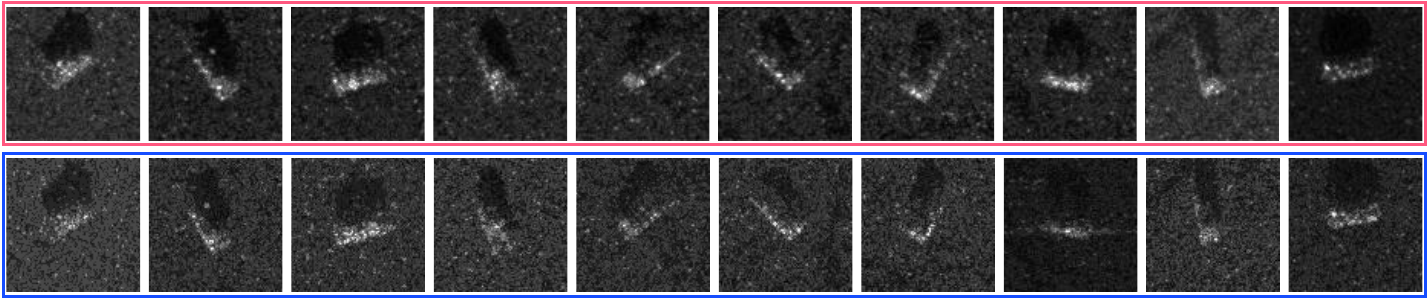}}
\subfigure[]{\label{1.2}\includegraphics[width=6in]{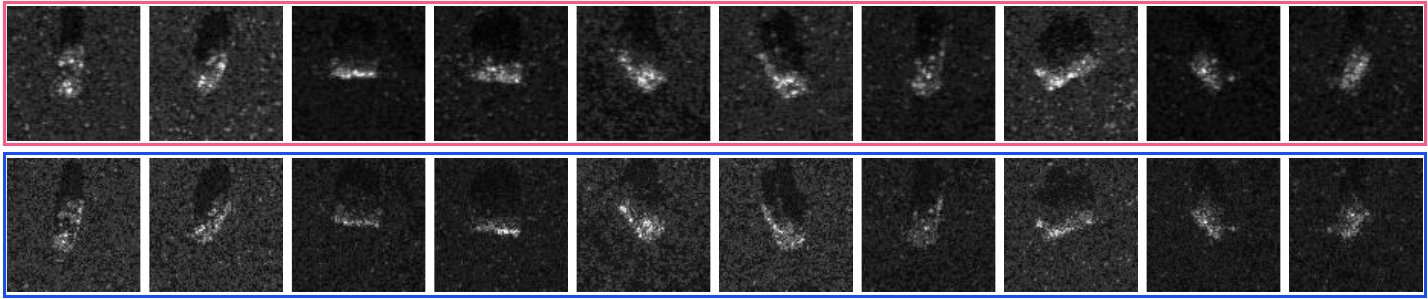}}
\subfigure[]{\label{1.3}\includegraphics[width=6in]{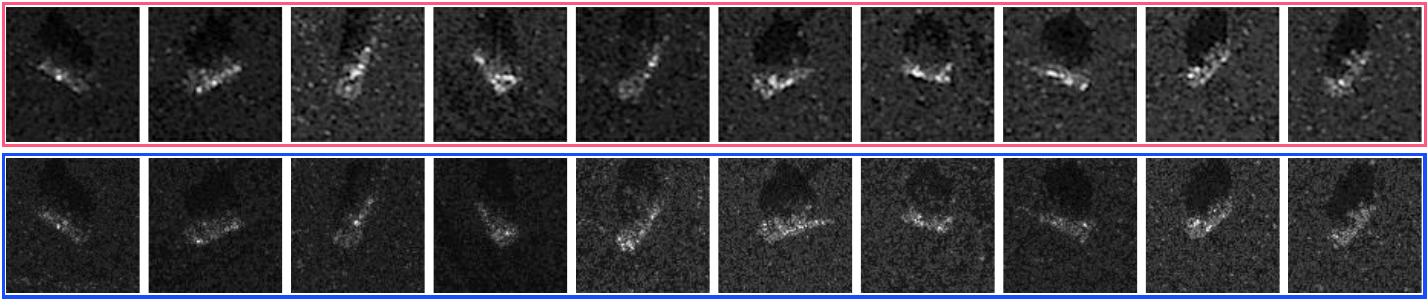}}
\subfigure[]{\label{1.4}\includegraphics[width=6in]{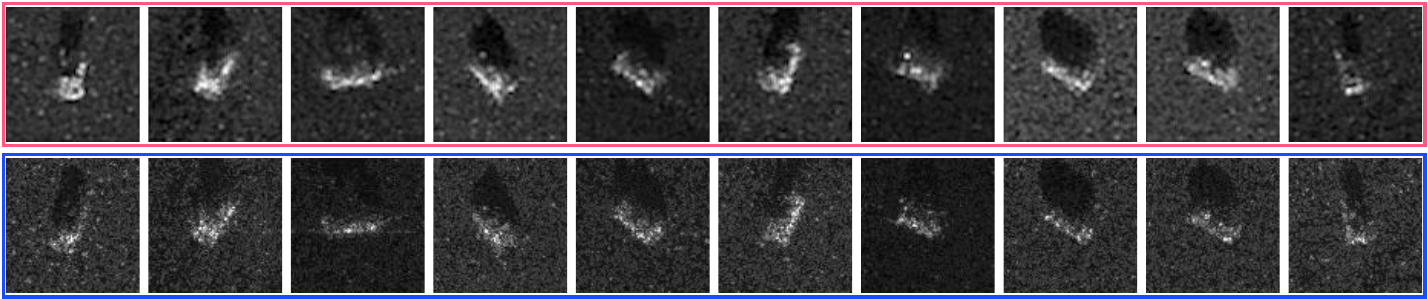}}\\
\end{center}
\caption{10 randomly chosen generated SAR images and real SAR images with close azimuths The red box is the generated SAR images and the blue is the real SAR images. (a) generated SAR images and corresponding real SAR images with 5\textdegree azimuth interval, (b) with 10\textdegree azimuth interval, (c) with 15\textdegree azimuth interval, and (d) with 20\textdegree azimuth interval.}
\label{figure5}
\end{figure*}

From Fig.\ref{figure5}, by comparing with the real SAR images with close azimuth, it is quite clear that the generated SAR images can acquire accurate geometric features and morphological structures in real SAR images. When the target type and azimuth are varying, the generated SAR images can still maintain high quality. Furthermore, through comparing the generated with the real from top to bottom, it is clear that the quality of the generated SAR images can still preserve enough similarity to the real despite the image details are fading a little when the azimuth interval is increasing. As a result, when the azimuth intervals in the dataset are increasing, the size of the dataset is declining, and the proposed azimuth-controllable SAR target image generation network shows the resilience to the smaller datasets and still generates stably accurate SAR images.

\begin{figure}[!htb]
\begin{center}
\subfigure[]{\label{1.1}\includegraphics[width=1.5in]{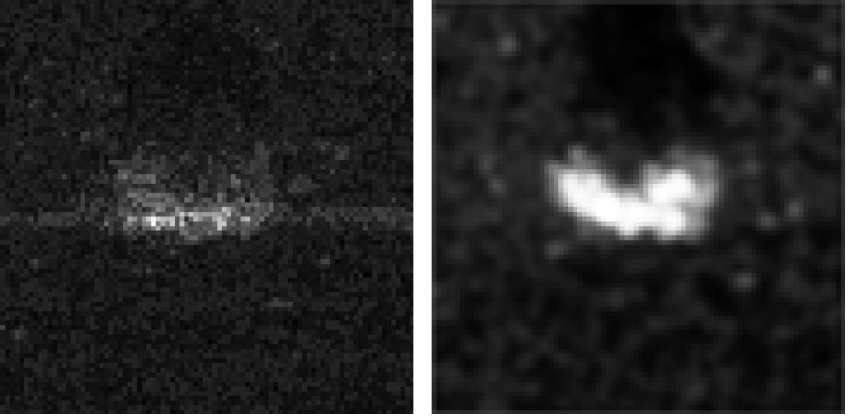}}
\subfigure[]{\label{1.2}\includegraphics[width=1.5in]{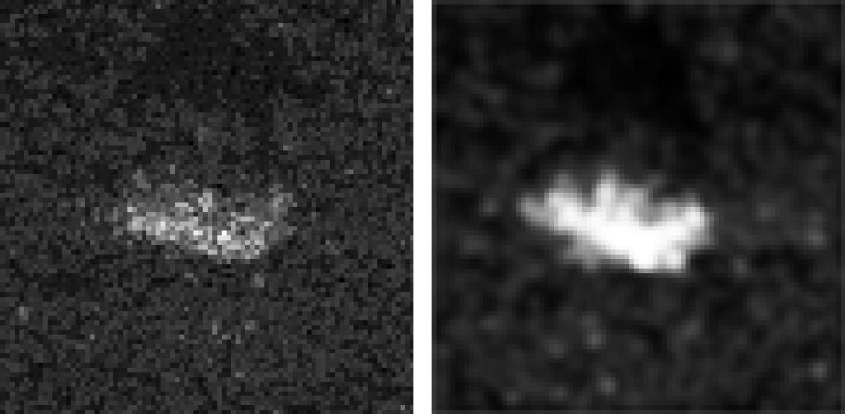}}
\subfigure[]{\label{1.3}\includegraphics[width=1.5in]{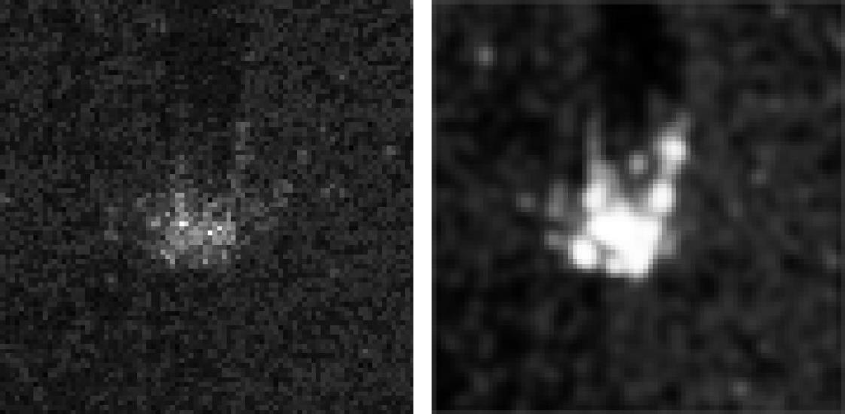}}
\subfigure[]{\label{1.4}\includegraphics[width=1.5in]{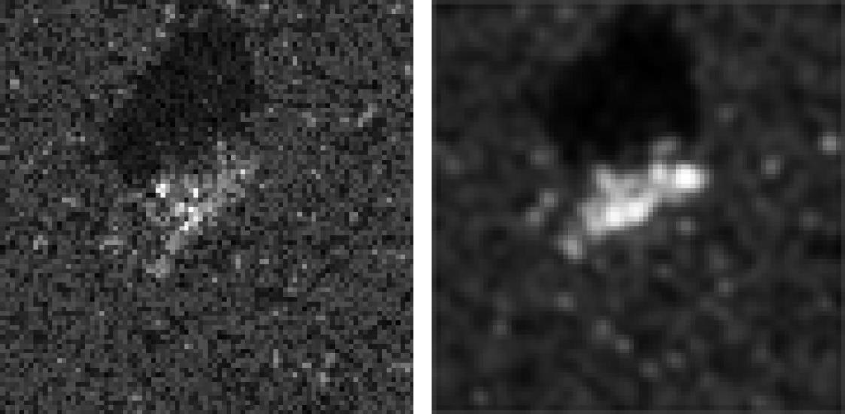}}\\
\end{center}
\caption{The generated SAR images and real SAR images under azimuth interval 30\textdegree.}\label{figure10}
\end{figure}

\begin{figure}[!htb]
\begin{center}
\subfigure[]{\label{1.1}\includegraphics[width=1.5in]{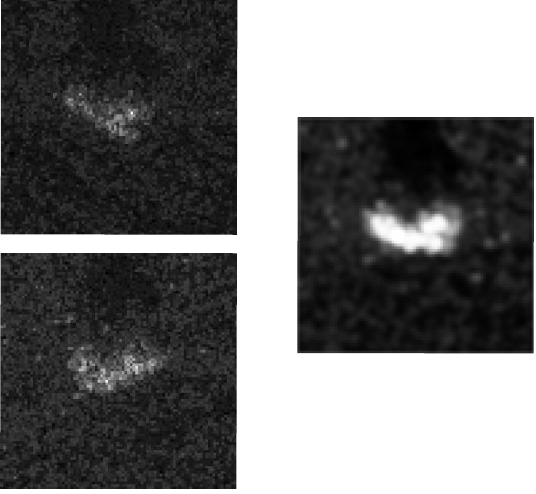}}
\subfigure[]{\label{1.2}\includegraphics[width=1.5in]{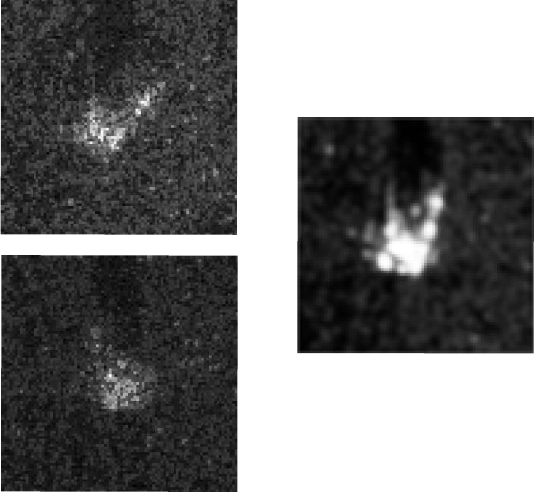}}
\subfigure[]{\label{1.3}\includegraphics[width=1.5in]{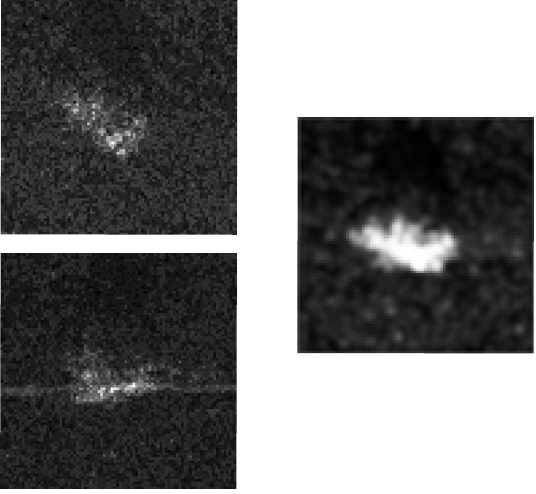}}
\subfigure[]{\label{1.4}\includegraphics[width=1.5in]{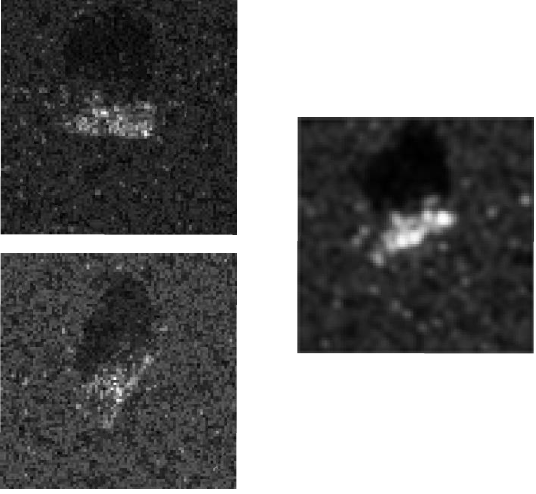}}\\
\end{center}
\caption{The generated SAR images and input SAR images of the generator under azimuth interval 30\textdegree.}\label{figure11}
\end{figure}

We had carried out some experiments about the border line of the training azimuth interval that can generate high-quality SAR images. From Fig.\ref{figure5}, when the azimuth interval is increasing from 5\textdegree to  20\textdegree, the quality of the generated SAR images is decreasing obviously. And when the azimuth interval is 20\textdegree, the generated SAR images start to obviously have the problem of over-smoothing. 
Moreover, we had generated SAR images with the azimuth interval 30\textdegree in Fig.\ref{figure10} and Fig.\ref{figure11}. The distributions of the generated SAR images are not similar to the real images, So, from the results of experiments, the border line should be between 20\textdegree to 30\textdegree.

In conclusion, the proposed azimuth-controllable SAR target image generation network can generate accurate SAR images with precise geometric features and morphological structures with the azimuth controllability, when the target type, azimuth, and the azimuth interval differ.

\subsubsection{Evaluation in Quantitative Similarity}

To evaluate the generated SAR images more objectively, three common metrics of image similarity are employed. The three metrics are mean square error (MSE) \cite{willmott2005advantages}, structural similarity (SSIM) \cite{wang2004image}, and mean structural similarity (MSSIM) \cite{kandadai2008audio}. MSE is a direct distance between the real SAR images and generated SAR images. SSIM focus on the whole images, and MSSIM more focus on the local details in images. They can be calculated as follows.

{
\setlength\abovedisplayskip{2.5pt}
\begin{flalign}
\label{rangereso1}
MSE = \frac{1}{{m \times n}}\sum\limits_{i = 1}^m {\sum\limits_{j = 1}^n {{{\left( {{\bf{x}}\left( {i,j} \right) - {\bf{y}}\left( {i,j} \right)} \right)}^2}} }
\end{flalign}}
where $\bf{y}$ denotes the generated SAR images and $\bf{x}$ denotes the real SAR images with close azimuth, $m$ and $n$ are the length and width of SAR images. The smaller MSE is, the higher the similarity is between real SAR images and generated SAR images.

{
\setlength\abovedisplayskip{2.5pt}
\begin{flalign}
\label{rangereso1}
SSIM = \frac{{\left( {2{\mu _x}{\mu _y} + {c_1}} \right)\left( {2{\sigma _{xy}} + {c_2}} \right)}}{{\left( {\mu _x^2 + \mu _y^2 + {c_1}} \right)\left( {\sigma _x^2 + \sigma _y^2 + {c_2}} \right)}}
\end{flalign}}
where ${{\mu }_{x}}$ and ${{\sigma }_{x}}$denotes the mean value and the standard deviation of the real images $x$, ${{c}_{1}}$ and ${{c}_{2}}$ are two constants related to the dynamic range of the pixel values. SSIM ranges from -1 to 1, where 1 indicated perfect similarity.

MSSIM divided the SAR images into $N$ blocks by sliding window, then calculate the weighted mean, variance, and covariance of all the blocks by ${{w}_{i,j}}$, and $\sum\limits_{i}{\sum\limits_{j}{{{w}_{i,j}}}}=1$, the SSIM of each block is obtained. Finally, the average value of all the SSIM values of all the blocks is set as MSSIM.

{
\setlength\abovedisplayskip{2.5pt}
\begin{flalign}
\label{rangereso1}
MSSIM = \frac{1}{N}\sum\limits_{k = 1}^N {SSIM({{\bf{x}}_k},{{\bf{y}}_k})}
\end{flalign}}
where ${{\bf{x}}_{k}}$ denotes the $kth$ block of a generated SAR image, ${{\bf{x}}_{k}}$ denotes the $kth$ block of a real SAR image. Same as SSIM, the higher is better. Before the calculation of the three metrics, the generated images and real images are normalized to the range $\left[ 0,255 \right]$.

\begin{table}[!htb]
\small
\centering
\begin{spacing}{1.4}
\caption{Quantitative results under increasing azimuth interval}\label{table3}
\begin{tabular}{p{1cm}<{\centering}p{1.2cm}<{\centering}p{1cm}<{\centering}p{1.2cm}<{\centering}p{2cm}<{\centering}}
\hline \hline
  & MSE & SSIM & MSSIM & Azimuth Error\\
\hline
5\textdegree&	0.00029	 & 0.73	& 0.9997 & 1.43 \\
\hline
10\textdegree& 0.00035	& 0.57	& 0.9996 & 2.88 \\
\hline
15\textdegree&	0.00052	& 0.52	& 0.9994 & 4.79 \\
\hline
20\textdegree&	0.00058	& 0.51	& 0.9993 & 7.25\\
\hline \hline
\end{tabular}
\end{spacing}
\end{table}

As shown in Table \ref{table3}, the quantitative results of the testing dataset of different azimuth intervals are presented from top to bottom. From Table \ref{table3}, it can be seen that as the azimuth interval is increasing, the value of MSE is still low, which means the generated images are quite similar to the real SAR images. Besides, when the azimuth interval is increasing, the value of SSIM is decreasing from 0.73, and the value of MSSIM are maintaining stable more than 0.99. In short, for the three metrics, the generated SAR images are of high quality and precise in visual similarity.

For better verifying the effectiveness of the method, we compared the generation images by our method with other excellent generation methods, such as WGAN \cite{arjovsky2017wasserstein}, DCGAN \cite{radford2015unsupervised} and SARGAN \cite{guo2017synthetic}. The comparison of MSE, SSIM and MSSIM is listed in Table \ref{table11}. The results summarized in Table \ref{table11} show that our model gets a better score than others. It is clear that the generated SAR images by our method are more similar to the real SAR images.

Through the visual and numerical presentation above, it is clear that the proposed azimuth-controllable SAR target image generation network has the capability of generating precise SAR images with high quality between the two adjacent azimuths with preserving the geometry and local details when the target type and azimuth vary. When the azimuth interval of the training/testing dataset is increasing and the size of the training/testing dataset is declining, the proposed azimuth-controllable SAR target image generation network can still generate precise SAR images based on the real SAR images with azimuth controllability.

To evaluate the applicative capability of the generated images further, the recognition experiment will be present under the declining training/testing dataset.

\begin{table}[!htb]
\small
\centering
\begin{spacing}{1.4}
\caption{Comparison of quantitative results with other generation methods}\label{table11}
\begin{tabular}{p{1cm}<{\centering}p{1.32cm}<{\centering}p{1.32cm}<{\centering}p{1.32cm}<{\centering}p{1.32cm}<{\centering}}
\hline \hline
  & OURS &   WGAN & DCGAN & SARGAN\\
\hline
MSE &	0.00029	 & 0.00042	 & 0.00044	& 0.00048 \\
\hline
SSIM & 0.73	& 0.71 & 0.68	& 0.61\\
\hline
MSSIM &	0.9997	&	0.9689 & 0.9542	& 0.9578 \\
\hline \hline
\end{tabular}
\end{spacing}
\end{table}

\subsection{Evaluation of SAR ATR Performance Improvements}

\begin{table*}[!t]
\renewcommand{\arraystretch}{1.3}
\caption{Entire dataset of training and testing for SOC}\label{table4}
\centering
\begin{tabular}{p{3.2cm}<{\centering}|p{3cm}<{\centering}|p{3cm}<{\centering}|p{3cm}<{\centering}|p{3cm}<{\centering}}
\hline \hline
  & \multicolumn{2}{c|}{Training} & \multicolumn{2}{c}{Testing}  \\
\hline
Class  & Depression & Number & Depression & Number  \\
\hline
BMP2-9563  & 17\textdegree & 116 & 15\textdegree & 196  \\
\hline
BTR70-c71 & 17\textdegree & 116 & 15\textdegree & 196  \\
\hline
T72-132 & 17\textdegree & 116 & 15\textdegree & 196  \\
\hline
BTR60-7532 & 17\textdegree & 128 & 15\textdegree & 195  \\
\hline
2S1-b01 & 17\textdegree & 149 & 15\textdegree & 274  \\
\hline
BRDM2-E71 & 17\textdegree & 149 & 15\textdegree & 274  \\
\hline
D7-92 & 17\textdegree & 149 & 15\textdegree & 274  \\
\hline
T62-A51 & 17\textdegree & 149 & 15\textdegree & 273  \\
\hline
ZIL131-E12& 17\textdegree & 149 & 15\textdegree & 274  \\
\hline
ZSU234-d08 & 17\textdegree & 149 & 15\textdegree & 274  \\
\hline \hline
\end{tabular}
\end{table*}

In this subsection, the improvement of the generated SAR images for the recognition performance will be evaluated under the standard operating condition (SOC) and extended operating condition (EOC) separately. SOC refers to that the serial numbers and target configurations of the train and test set are the same, but with different aspects and depression angles. EOC includes three extended operating conditions: depression variant, configuration variant, and version variant. The comparison will be presented in the aspect of the recognition performance and the limited size of the training dataset.

We employed the recognition network proposed in \cite{chen2016target} and it is denoted as A-ConvNets. This network is composed of five convolution layers, three max-pooling layers. And it has achieved high performance of recognition under the full MSTAR dataset. Therefore, it is employed to evaluate the performance of recognition without and with the generated SAR image by our proposed network.

\subsubsection{Improvement of Recognition Results under SOC}

In the SOC experiment, the training dataset is captured at 17\textdegree depression angle, the testing is 15\textdegree depression angle. The entire dataset of the training and testing has been listed in Table \ref{table4}. And under different azimuth intervals, the size of the dataset of training and testing will change. Therefore, for immediately apparent presentation, the summary of the training and testing dataset for the recognition performance is denoted as the azimuth interval listed in Table \ref{table5}. For enough training data of CNN, the data listed in Table \ref{table5} is augmented 10 times by randomly sampling ten $88\times88$ SAR image chips from one original $128\times128$ SAR image, which ensures the target complete \cite{chen2016target}.

\begin{figure*}[!htb]
\begin{center}
\subfigure[]{\label{1.1}\includegraphics[width=2.5in]{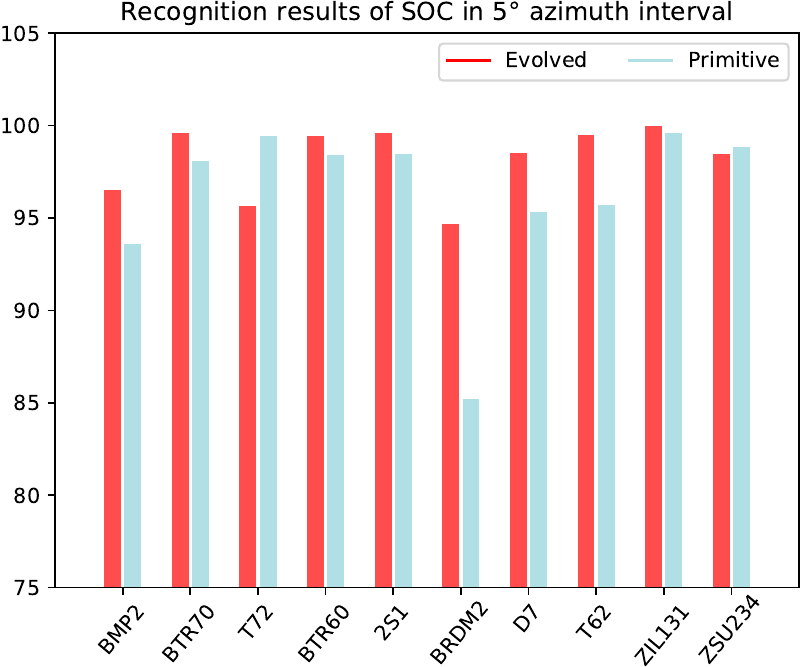}}
\subfigure[]{\label{1.2}\includegraphics[width=2.5in]{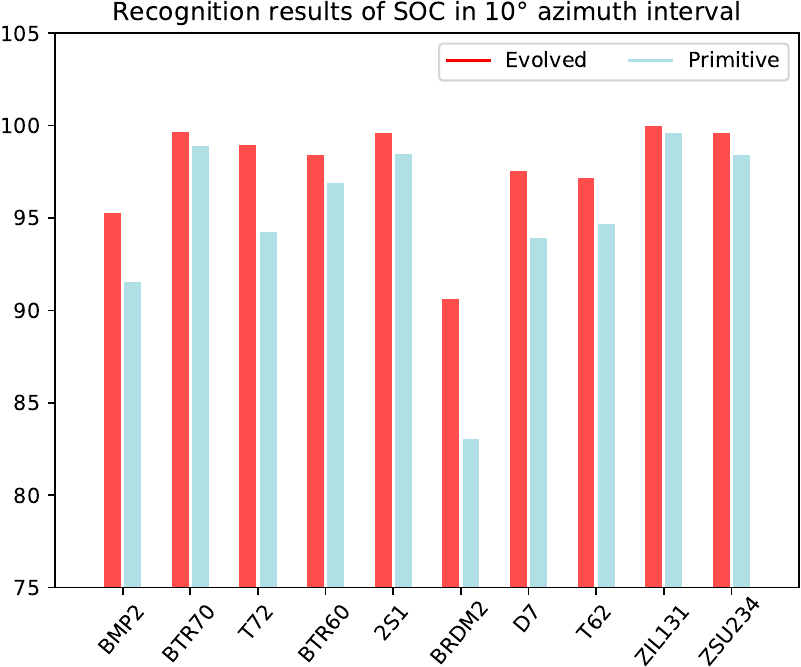}}
\subfigure[]{\label{1.3}\includegraphics[width=2.5in]{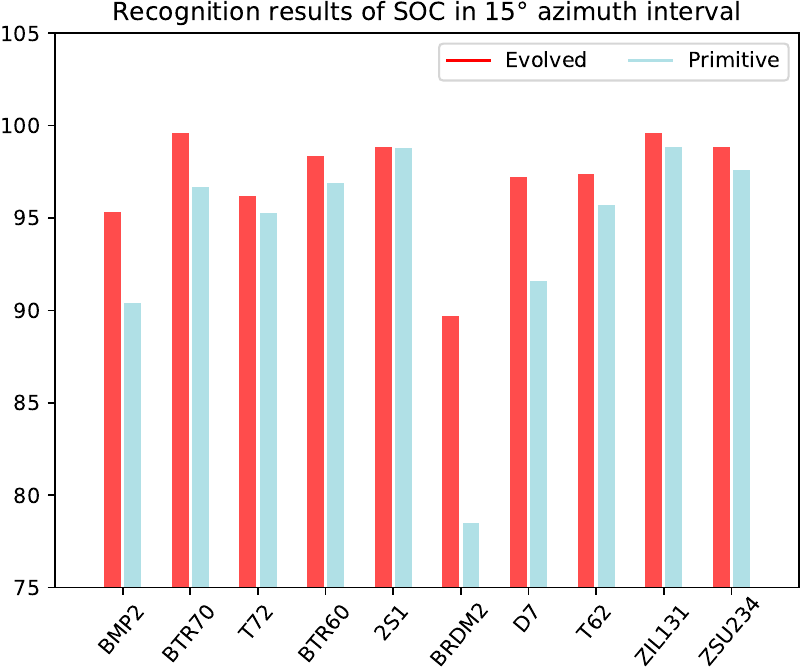}}
\subfigure[]{\label{1.4}\includegraphics[width=2.5in]{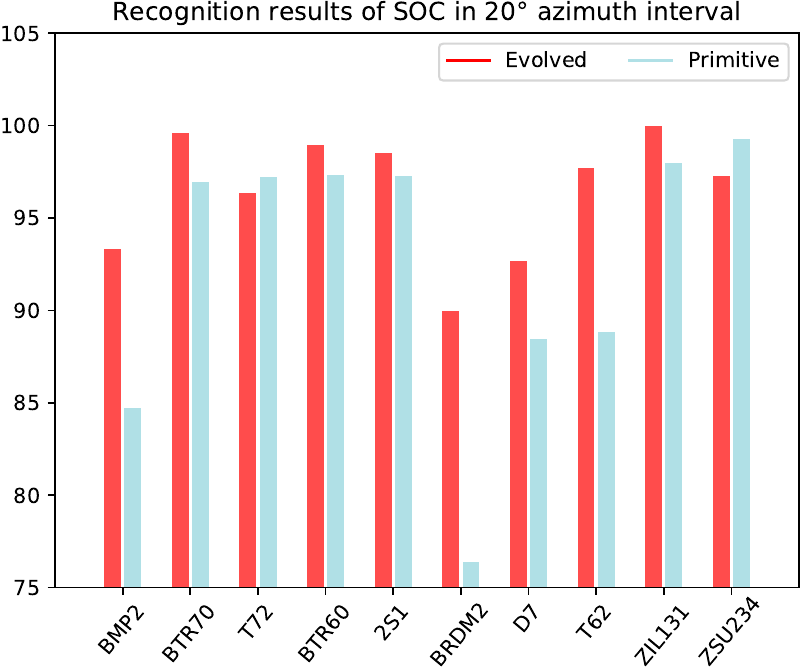}}\\
\end{center}
\caption{Primitive and evolved recognition results of different azimuth intervals under SOC.}\label{figure6}
\end{figure*}

The primitive recognition results are presented in Fig.\ref{figure6} with the blue bar, which show the training dataset is only the declined dataset without the generated SAR images. The evolved recognition results, whose training dataset is stacked by the declined dataset and the generated SAR images, are listed in Fig.\ref{figure6} with the red bar. And the targets of the histograms in Fig.\ref{figure6} are sequenced as BMP2-9563, BTR70-c71, T72-132, BTR60-7532, 2S1-b01, BRDM2-E71, D7-92, T62-A51, ZIL131-E12, and ZSU234-d08.

\begin{table*}[!htb]
\small
\centering
\begin{spacing}{1.19}
\caption{Number of training and testing dataset for the evaluation of the recognition performance under SOC before the data augmentation}\label{table5}
\begin{tabular}{p{3.5cm}<{\centering}p{4.5cm}<{\centering}p{4.5cm}<{\centering}p{3.2cm}<{\centering}}
\hline \hline
\multirow{2}*{Azimuth interval} & \multicolumn{2}{c}{Training} & \multirow{2}*{Testing}  \\
~ & Without generated SAR images  & With generated SAR images & ~ \\
\hline
5\textdegree&	550 &	1099	& 2425  \\
\hline
10\textdegree&  308 &	615 &	2425 \\
\hline
15\textdegree&	214	 &427	& 2425 \\
\hline
20\textdegree&	167 & 	333	 & 2425 \\
\hline \hline
\end{tabular}
\end{spacing}
\end{table*}

As (a) in Fig.\ref{figure6}, the overall recognition rates are improved from 96.22\% to 98.22\% in the azimuth interval 5\textdegree. For most of the target types, the recognition rates are improved at least 2.00-3.00\%, especially for I2S1 85.23\% to 94.66\%. And as for (b) in Fig.\ref{figure6}, the overall recognition rate is improved from 95.05\% to 97.74\%. Same as the azimuth interval 5\textdegree, the recognition rates of most of the target types are improved obviously. Besides, the recognition rates of BMP2 and I2S1 are improved from 91.53\% to 95.29\% and from 83.02\% to 90.64\%, which greatly limits the overall recognition rates. Then as (c) in Fig.\ref{figure6}, the overall recognition rate is improved from 93.88\% to 97.14\%. Although the recognition rates of the BMP2, I2S1, and T62 is not high enough, the others can still get an obvious promotion. Finally, as (d) in Fig.\ref{figure6}, the overall recognition rate is improved from 92.53\% to 96.44\%. The primitive recognition rates of BMP2, T62, and T72 are lower than 90.00\% and the recognition rates of I2S1 are lower than 80.00\%. After employing the generated SAR images, the recognition rates of BMP2, T62, T72, and I2S1 are improved to 93.33\%, 92.68\%, 97.71\%, and 89.96\% separately, whose improvements is 8.59\%, 4.25\%, 13.56\%, and 8.88\%. In conclusion, the comparison and improvement have demonstrated that through the employment of the generated SAR images, under different azimuth intervals, the stable distinguishable features for the recognition increase, and the recognition rates performance is improved obviously.

As the blue bar of the primitive results in Fig.\ref{figure6}, the overall recognition rate is decreasing from 96.22\% to 92.53\% gradually. Therefore, it is clear that the recognition rate is decreasing when the azimuth interval is increasing and the size of the training dataset is declining. From the different improvement in different azimuth interval, 2.00\% at 5\textdegree, 2.69\% at 10\textdegree, 3.26\% at 15\textdegree, and 3.91\% at 20\textdegree, it can be summarized that the recognition rate can be improved more with the more recognition information provided by the generated SAR images when the size of the dataset is declining. From all the comparison and improvement above, it can be demonstrated that the recognition results can be promoted through the employment of the generated images in the SOC, and the stable distinguishable features for the recognition is increased among different targets. It can demonstrate the superiority of the proposed azimuth-controllable SAR target image generation network.

\begin{figure}[!htb]
    \centering
        \includegraphics[width=0.5\textwidth]{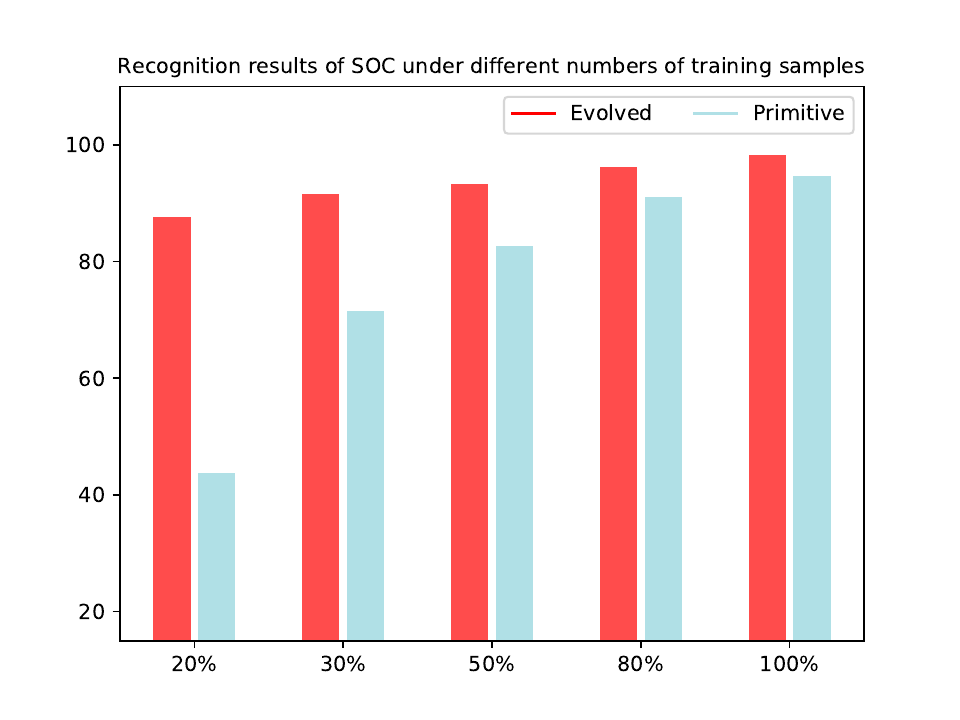}
    \caption{Recognition results of SOC under different numbers of training samples.}\label{figure12}
\end{figure}

We have added some comparison experiments with different numbers of real images. The fake SAR images are generated in the 5\textdegree azimuth interval and the real images are chosen from these real images which were used to generate the fake SAR images. The recognition performances are as follow.
In this Fig 1, the red bar denotes the training sample contains real SAR images and generated images, called evolved. And the powder-blue bar denotes the training sample only contains real SAR images, called primitive. As shown as X-axis in Fig.\ref{figure12}, we set the training samples as 80.00\%, 50.00\%, 30.00\% and 20.00\%. To keep the number of training samples the same between evolved and primitive, the evolved training samples are augmented 10 times, and the primitive training samples are augmented 20 times.
From the results in Fig.\ref{figure12} , the evolved recognition performances of SOC are obviously higher than the results of only real SAR images. From 100.00\% to 20.00\% training samples, the recognition ratios of primitive are decreasing prominently, but the evolved performances are robust against the decreased training samples. At the 20.00\% situation, the primitive result seems to fail to recognize, but the evolved result still stays around 89.00\%. It is clear that the generated SAR images can promote higher recognition performance than only real SAR images of different numbers of training samples.

\subsubsection{Improvement of Recognition Results under EOC}

\begin{table}[!htb]
\small
\centering
\begin{spacing}{1.19}
\caption{Number of training datasets under EOC-D before the data augmentation}\label{table6}
\begin{tabular}{p{2.2cm}<{\centering}p{2.5cm}<{\centering}p{2.5cm}<{\centering}}
\hline \hline
\multirow{3}{*}{Azimuth interval} & \multicolumn{2}{c}{Training} \\
 & Without  & With \\
 & generated & generated \\
 & SAR images &  SAR images  \\
\hline
5\textdegree&	226	& 451  \\
\hline
10\textdegree& 127	& 253 \\
\hline
15\textdegree&	88	& 175 \\
\hline
20\textdegree&	69	& 137\\
\hline \hline
\end{tabular}
\end{spacing}
\end{table}

In the practical application of SAR ATR, there are many limitations in the recognition operation, such as the variances of the depression angle and target type. Therefore, it is a quite important aspect of evaluation in EOC. In this section, the performance of the SAR images generated by the proposed azimuth-controllable SAR target image generation network will be assessed in the variances of the depression angle, target configuration, and version, which is denoted as EOC-D, EOC-C, and EOC-V, respectively.

\begin{table}[!htb]
\small
\centering
\begin{spacing}{1.4}
\caption{Number of testing images for EOC-D}\label{table7}
\begin{tabular}{p{3cm}<{\centering}p{2.1cm}<{\centering}p{2.1cm}<{\centering}}
\hline \hline
Class  & Depression & Number   \\
\hline
2S1   & 30\textdegree & 288 \\
\hline
BRDM2  & 30\textdegree & 287 \\
\hline
T72  & 30\textdegree & 288 \\
\hline
ZSU234  & 30\textdegree & 288 \\
\hline \hline
\end{tabular}
\end{spacing}
\end{table}

\begin{figure*}[!htb]
\begin{center}
\subfigure[]{\label{1.1}\includegraphics[width=3in]{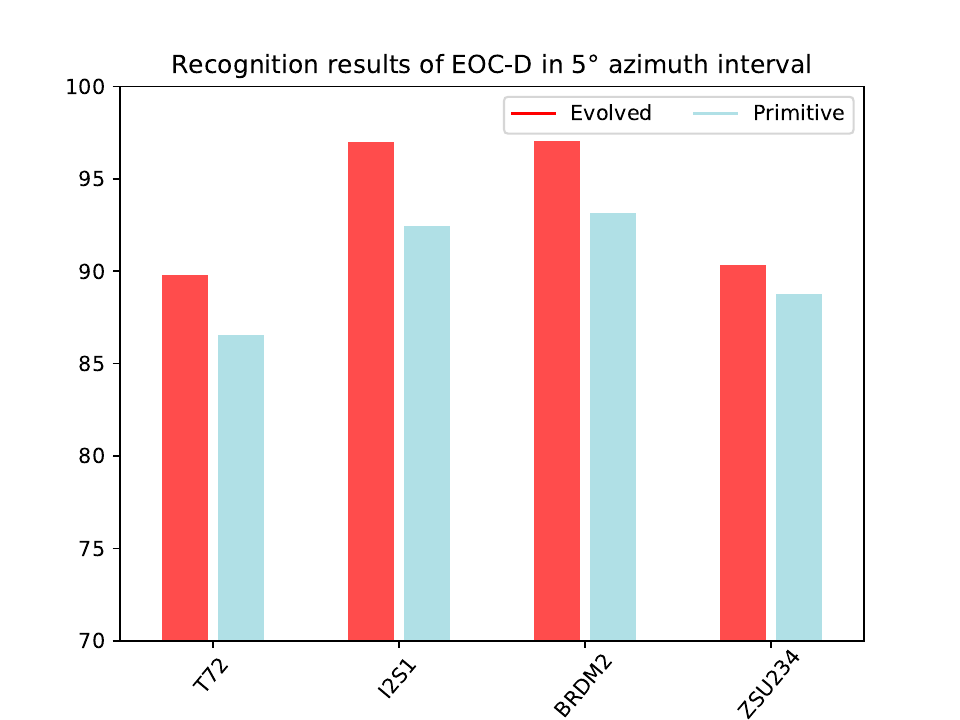}}
\subfigure[]{\label{1.2}\includegraphics[width=3in]{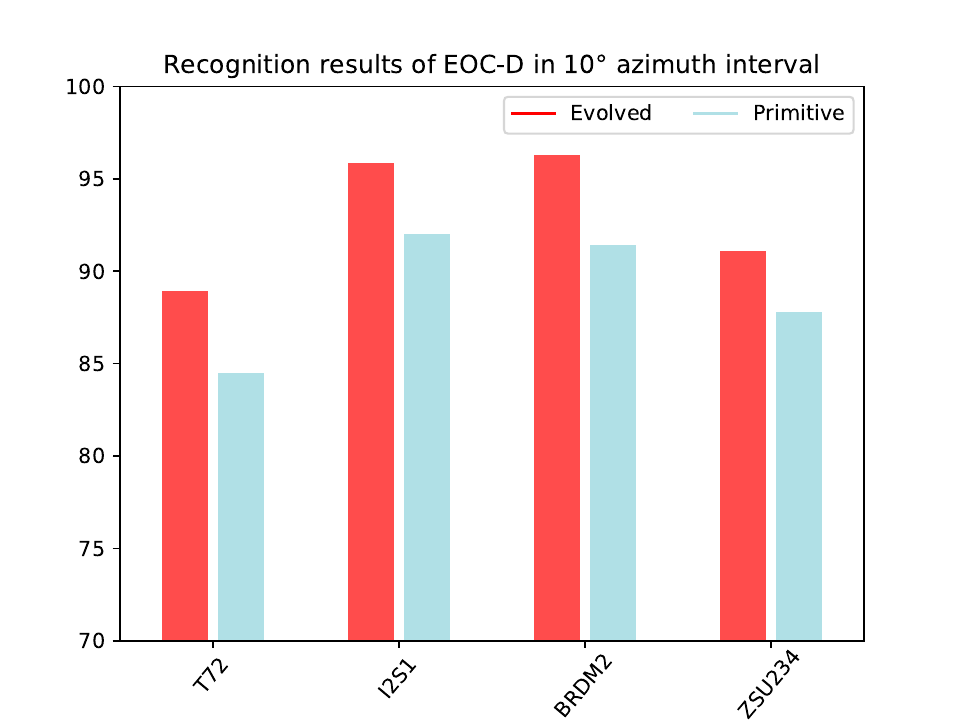}}\\
\subfigure[]{\label{1.3}\includegraphics[width=3in]{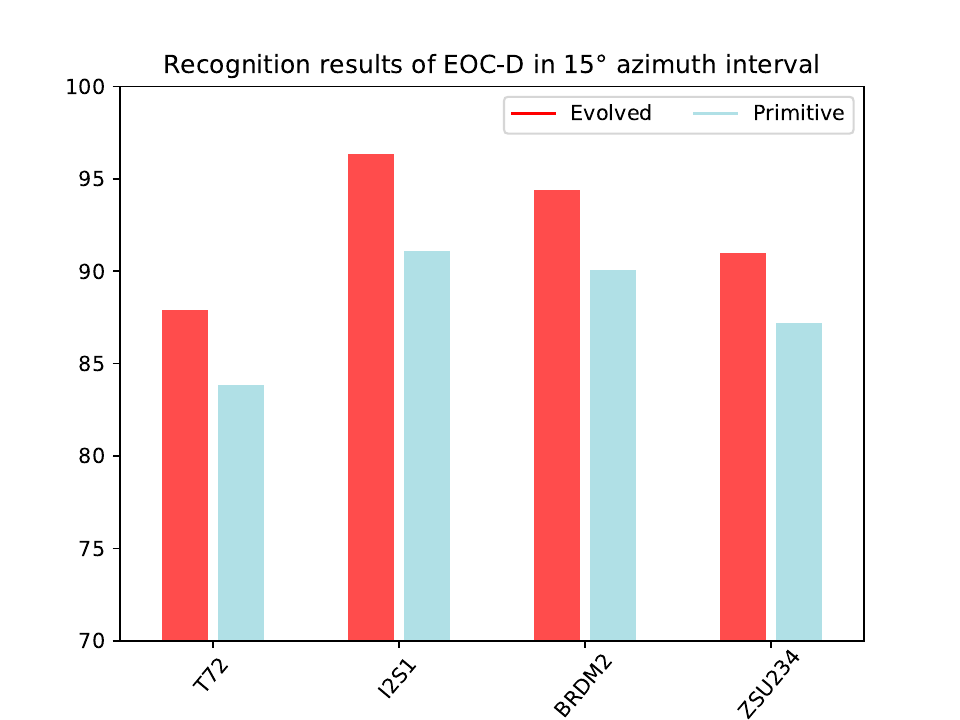}}
\subfigure[]{\label{1.4}\includegraphics[width=3in]{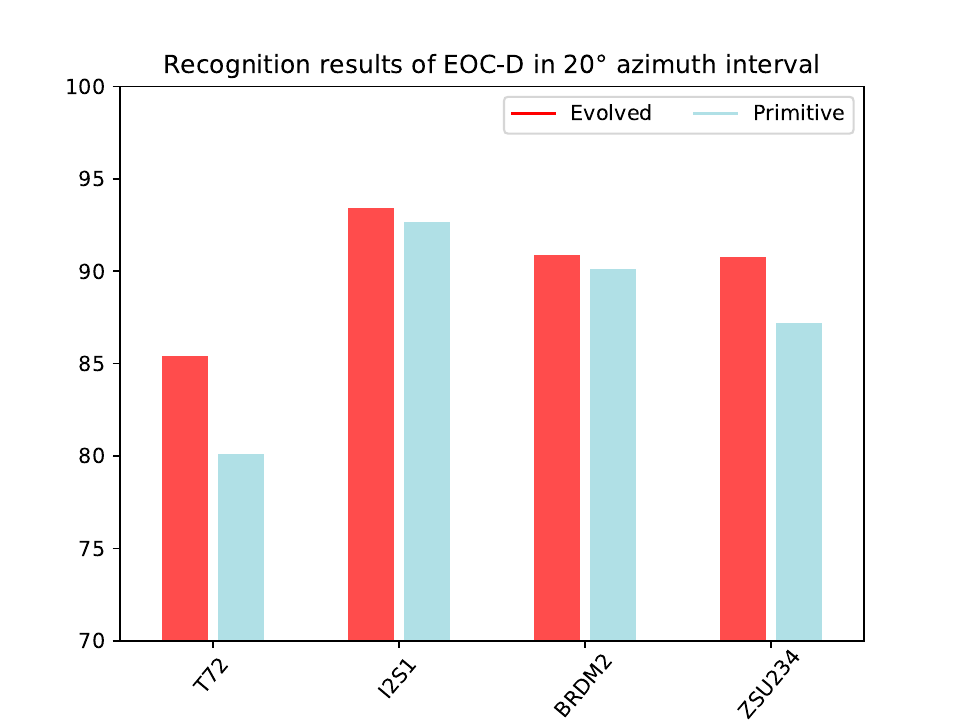}}\\
\end{center}
\caption{Primitive and evolved recognition results of different azimuth intervals under EOC-D.}\label{figure7}
\end{figure*}

The variance of the depression angle can extremely aggravate the performance of the recognition. Firstly, the performance of the generated SAR images will be assessed. Limited by that MSTAR dataset only contains four targets (2S1, BRDM-2, T-72, and ZSU-234) at 30\textdegree depression angle which are set as the testing dataset, the training dataset is the corresponding SAR images at 17\textdegree depression angle listed in Table \ref{table4}. And the declining training dataset is denoted as the azimuth interval listed in Table \ref{table6} and augmented by the same method as in SOC. The summary of the testing dataset is listed in Table \ref{table7} and the recognition performances of EOC-D under different azimuth intervals are listed in Fig.\ref{figure7}.

\begin{table}[!htb]
\small
\centering
\begin{spacing}{1.19}
\caption{Number of training datasets under EOC-C and EOC-V before the data augmentation}\label{table8}
\begin{tabular}{p{2.2cm}<{\centering}p{2.5cm}<{\centering}p{2.5cm}<{\centering}}
\hline \hline
\multirow{3}{*}{Azimuth interval} & \multicolumn{2}{c}{Training} \\
 & Without  & With \\
 & generated & generated \\
 & SAR images &  SAR images  \\
\hline
5\textdegree&	200 &	399  \\
\hline
10\textdegree& 112 &	223 \\
\hline
15\textdegree&	78	& 155 \\
\hline
20\textdegree&	61 &	121\\
\hline \hline
\end{tabular}
\end{spacing}
\end{table}

In Fig.\ref{figure7}, the primitive recognition rates of EOC-D are decreasing gradually from 90.22\% to 87.52\%, when the azimuth interval is increasing from 5\textdegree to 20\textdegree. Besides, the recognition rates are improved by 2.00-5.00\% from the primitive to the evolved. As for the recognition performance of each target type, the recognition rates of T72\_sn\_132 are the main limitation of the overall recognition performance, which are the lowest rates in the four types. And by comparing the primitive and evolved results of I2S1 and BRDM\_2 under different azimuth intervals, it is clear that the recognition performances of I2S1 and BRDM\_2 can benefit most from the generated SAR images with the resistance to the increasing azimuth interval. The recognition performance of ZSU\_234 is the most resistant to the increasing azimuth interval and unfortunately the generated SAR images. In short, the generated SAR images can be beneficial for the overall recognition performances of EOC-D under increasing azimuth interval by mainly improving the recognition of T72\_sn\_132, I2S1, and BRDM\_2.

\begin{table}[!htb]
\small
\centering
\begin{spacing}{1.4}
\caption{Number of testing images for EOC-C}\label{table9}
\begin{tabular}{p{1.9cm}<{\centering}p{1.6cm}<{\centering}p{1.6cm}<{\centering}p{1.6cm}<{\centering}}
\hline \hline
Class & Serial No. &  Depression & Number \\
\hline
\multirow{2}{*}{BMP2}
 &9566 & 15\textdegree,17\textdegree & 428  \\
\cline{2-4}
 &C21 & 15\textdegree,17\textdegree & 429 \\
\hline
\multirow{5}{*}{T72}
 &812& 15\textdegree,17\textdegree & 426 \\
\cline{2-4}
 &A04 & 15\textdegree,17\textdegree & 573 \\
\cline{2-4}
 &A05 & 15\textdegree,17\textdegree & 573 \\
\cline{2-4}
 &A07 & 15\textdegree,17\textdegree & 573 \\
\cline{2-4}
 &A10 & 15\textdegree,17\textdegree & 567 \\
\hline \hline
\end{tabular}
\end{spacing}
\end{table}

\begin{table}[!htb]
\small
\centering
\begin{spacing}{1.4}
\caption{Number of testing images for EOC-V}\label{table10}
\begin{tabular}{p{1.9cm}<{\centering}p{1.6cm}<{\centering}p{1.6cm}<{\centering}p{1.6cm}<{\centering}}
\hline \hline
Class & Serial No. &  Depression & Number \\
\hline
\multirow{5}{*}{T72}
 &S7 & 15\textdegree,17\textdegree & 419  \\
\cline{2-4}
 &A32 & 15\textdegree,17\textdegree & 572 \\
\cline{2-4}
 &A62 & 15\textdegree,17\textdegree & 573 \\
\cline{2-4}
 &A63 & 15\textdegree,17\textdegree & 573 \\
\cline{2-4}
 &A64 & 15\textdegree,17\textdegree & 573 \\
\hline \hline
\end{tabular}
\end{spacing}
\end{table}

The recognition performances at the variance of target configuration and version (EOC-C and EOC-V) are also evaluated. The training datasets for EOC-C and EOC-V include four targets(BMP-2, BRDM-2, BTR-70, and T-72) at a 17\textdegree depression angle listed in Table \ref{table4}. The numbers of the training data of the four targets are listed in Table \ref{table8} augmented by the same method as in SOC. The testing datasets for EOC-C and EOC-V are listed in Table \ref{table9} and Table \ref{table10}.

\begin{figure*}[!htb]
\begin{center}
\subfigure[]{\label{1.1}\includegraphics[width=3in]{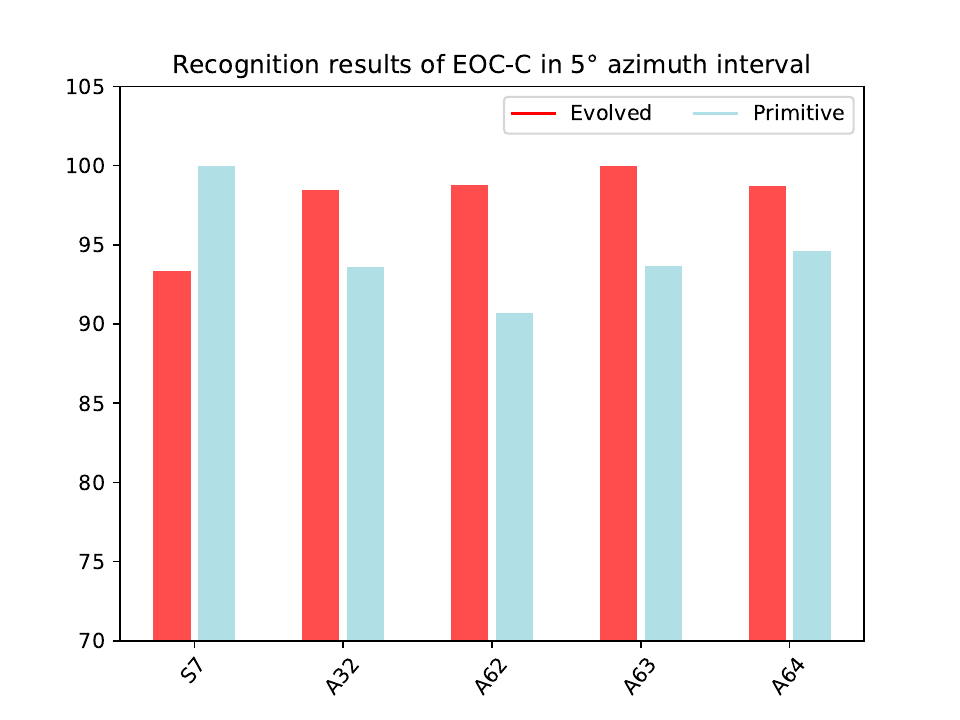}}
\subfigure[]{\label{1.2}\includegraphics[width=3in]{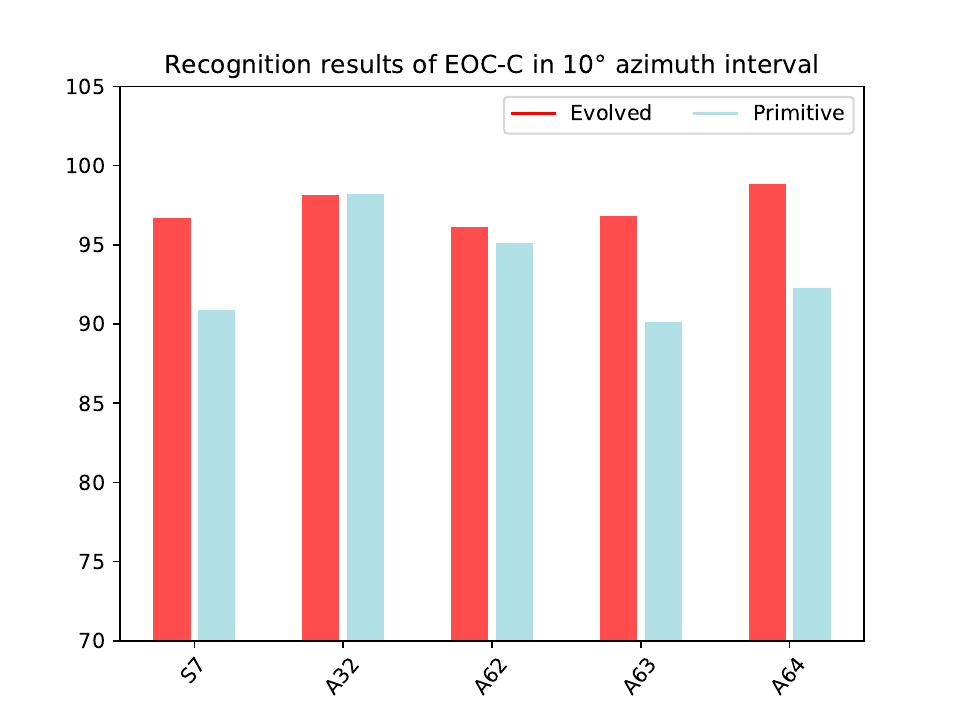}}
\subfigure[]{\label{1.3}\includegraphics[width=3in]{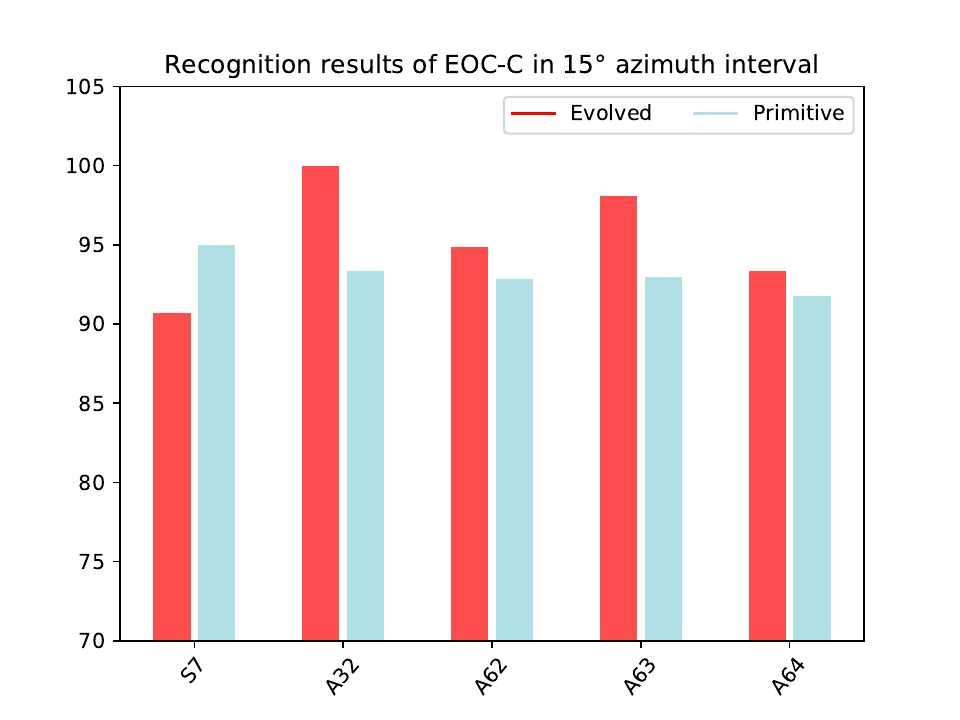}}
\subfigure[]{\label{1.4}\includegraphics[width=3in]{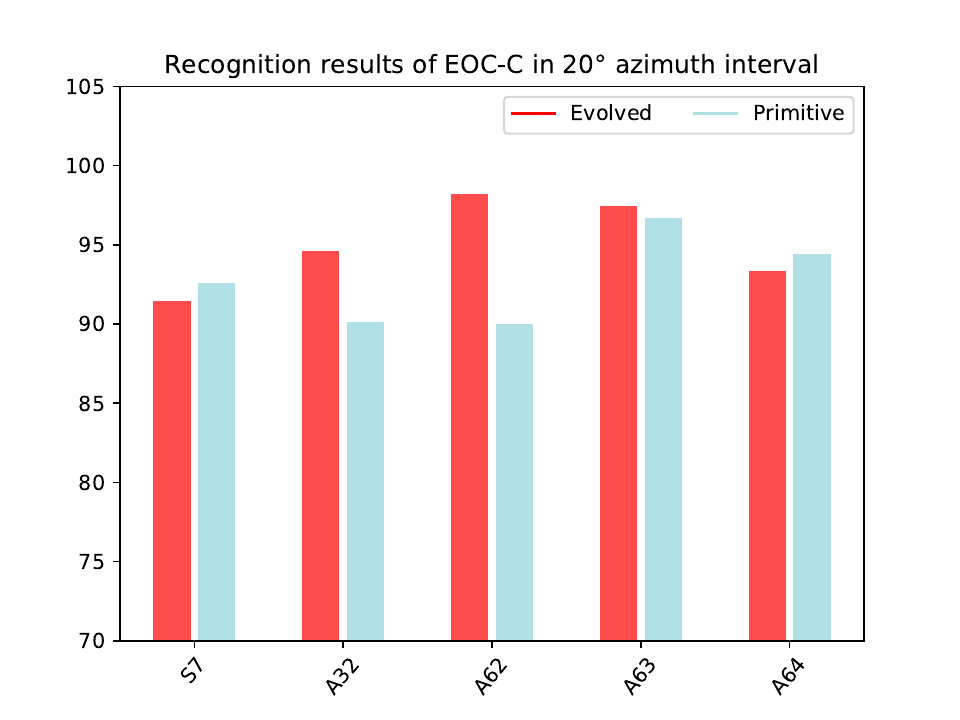}}\\
\end{center}
\caption{Primitive and evolved recognition results of different azimuth intervals under EOC-C.}\label{figure8}
\end{figure*}

From Table \ref{table9}, two different serial types of BMP2 and five different serial types of T72 captured at 17\textdegree and 15\textdegree depression angles are employed to evaluate the recognition performance under the EOC of the target configuration varieties, EOC-C. From Table \ref{table10}, there are four different serial types of T72 in the testing dataset captured at 17\textdegree and 15\textdegree depression angle and utilized to evaluate the recognition performance under the EOC of the target version varieties, EOC-V.

\begin{figure*}[!htb]
\begin{center}
\subfigure[]{\label{1.1}\includegraphics[width=3in]{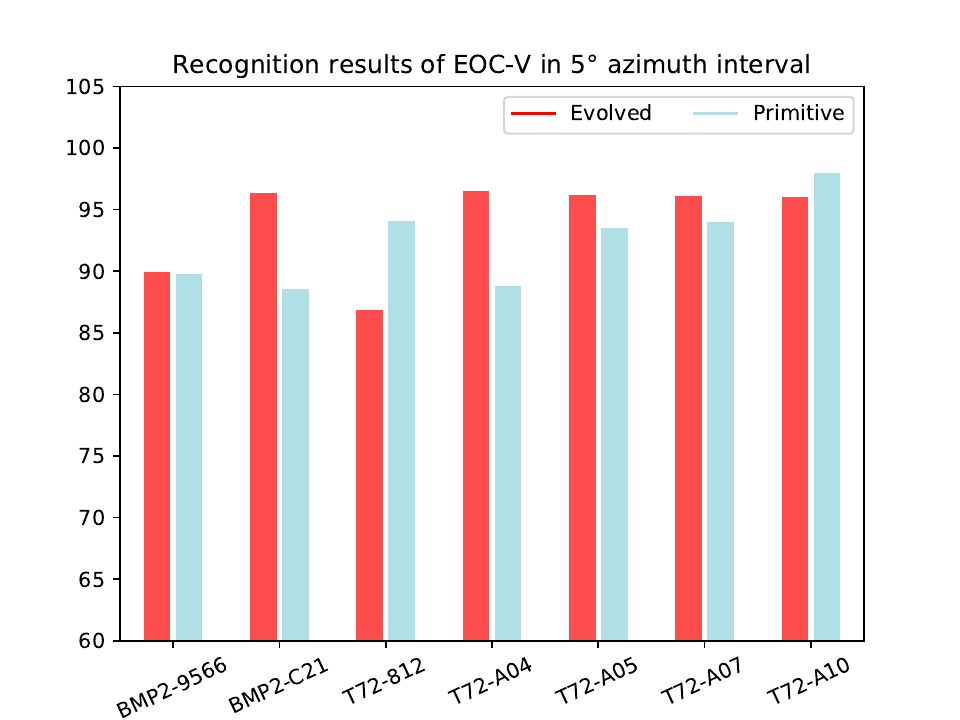}}
\subfigure[]{\label{1.2}\includegraphics[width=3in]{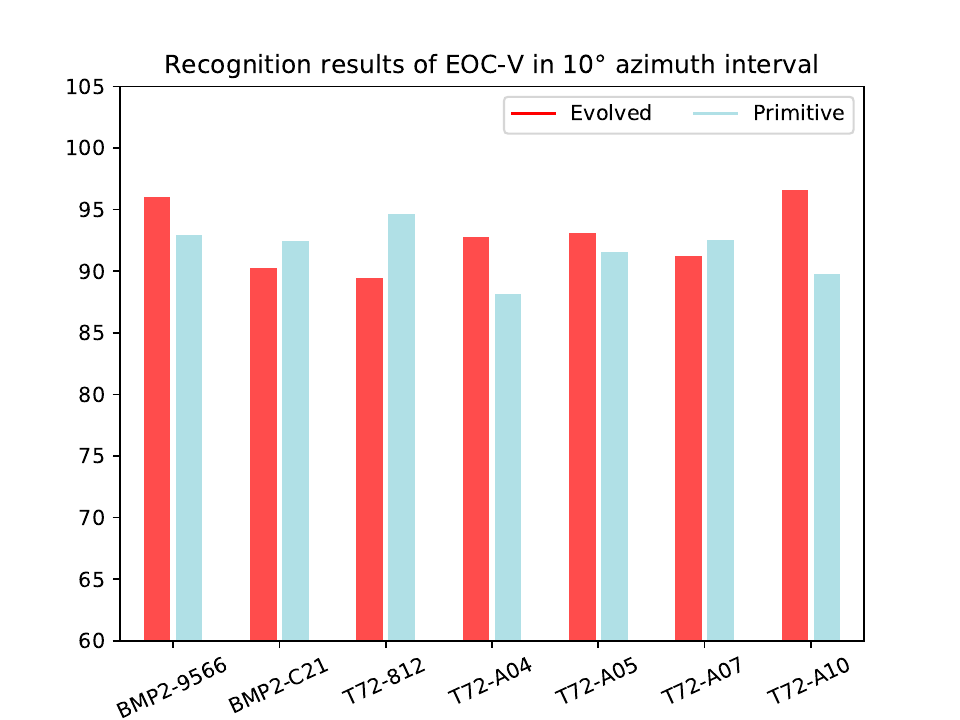}}
\subfigure[]{\label{1.3}\includegraphics[width=3in]{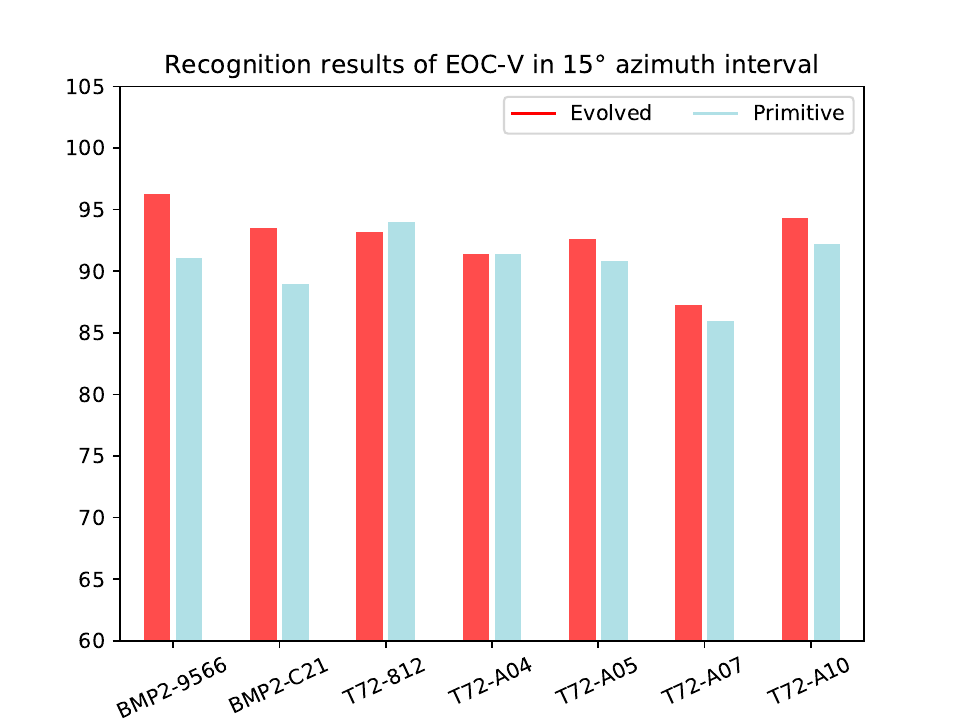}}
\subfigure[]{\label{1.4}\includegraphics[width=3in]{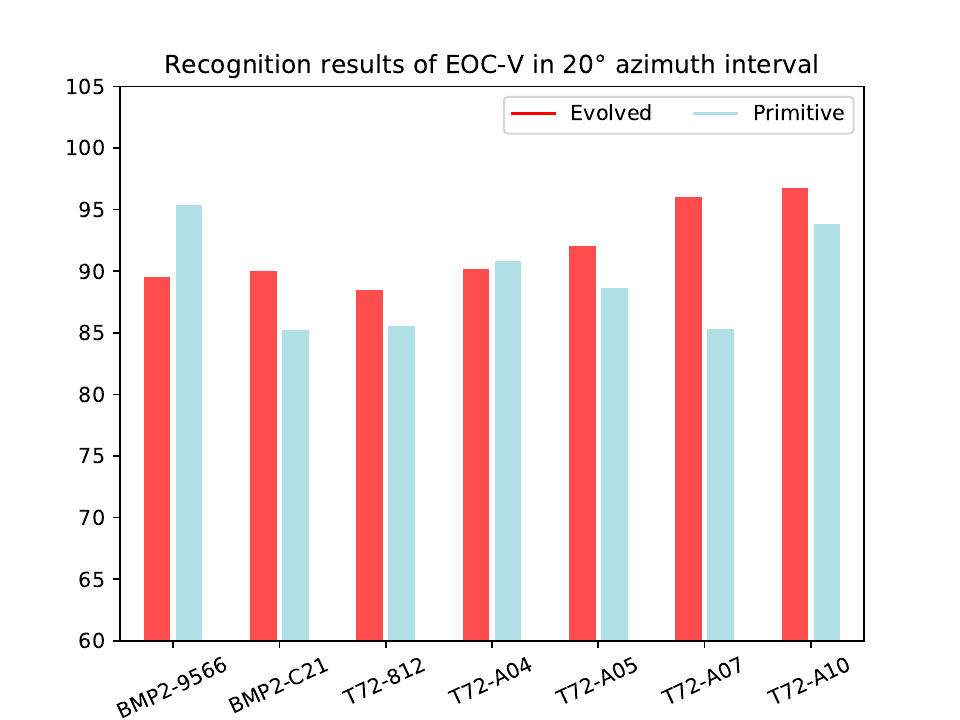}}\\
\end{center}
\caption{Primitive and evolved recognition results of different azimuth intervals under EOC-V.}\label{figure9}
\end{figure*}

The recognition performances of EOC-C under different azimuth intervals are listed in Fig.\ref{figure8}. And The recognition performances of EOC-V under different azimuth intervals are listed in Fig.\ref{figure9}.

By analyzing and comparing the performance of EOC-C under different azimuth intervals with crosswise and lengthwise in Fig.\ref{figure8}, it is clear that the evolved performance is slightly improved from the primitives. Although the performance is improved little when the azimuth interval is small as 5\textdegree or 10\textdegree, it can be improved around 3.00\% when the azimuth interval increases to 15\textdegree or 20\textdegree. In short, the generated SAR images can still be useful for the recognition of EOC-C.

From Fig.\ref{figure9}, it can demonstrate that the overall recognition rates of EOC-V under different azimuth intervals can be promoted 2.00-4.00\% by the employment of the generated SAR images. As the azimuth interval is increasing, the improving capability of the overall recognition rates is decreasing from 4.00\% to 2.00\%, which can result from the sensitiveness to the azimuth interval. Although the improvements of recognition are limited under high azimuth interval, it is obvious that the recognition rates are improved from 94.20\% to 98.07\% at 5\textdegree interval and 93.45\% to 97.36\% at 10\textdegree interval. In conclusion, the generated SAR images are meaningful for EOC-V.

From the four experiment results of SOC, EOC-D, EOC-C, and EOC-V, with superior recognition performance, the generated SAR images are beneficial for all the recognition performance. It demonstrates that the proposed azimuth-controllable SAR target image generation method has the capability of acquiring the precise distribution of SAR images in continuous azimuth and reconstructing the images from two adjacent images, which has great prospects in the applications of SAR ATR.

\subsection{Comparison with Other Augmentation Methods}
In this section, the proposed method will be compared with other augmentation methods in recognition. DNN1 \cite{dnn1}, DNN2 \cite{dnn2} and CNN+matrix \cite{cnnMetrix} used simple augmentation methods, like crop, rotate, and shift. WGAN-GP focuses on image data augmentation to generate new samples for SAR ATR \cite{wgangp}. DCGAN is employed to reduce the negative impact of the incorrectly labeled samples in SAR ATR \cite{radford2015unsupervised}. MGAN is for semi-supervised SAR ATR and aims to improve the performance of SAR ATR under a limited training dataset \cite{mgan}. SSDTL employs a variety of unlabeled samples for training a GAN \cite{ssdtl}. IGAN achieves semi-supervised generation and recognition simultaneously \cite{igan}. DNN2(PoseSy) means the recognition performance with the augmentation of pose synthesis, Multiscale \cite{9381993} employs randomly rotating and flipping, Weakly \cite{2021Weakly} employs randomly rotating in the recognition process.

\begin{table}[!htb]
\begin{center}
\caption{{{Recognition performance for various methods. }}}
\label{recognition}
\resizebox{\linewidth}{!}{
\begin{tabular}{ccccc}
\hline \hline
Method  & Around 200 & Around 350 & 400-800    & Above 800   \\
\hline
DNN1         & 77.86(200) & 86.98(400) & 93.04(800) & 95.54(all)  \\
DNN2         & 79.63(200) & 87.73(400) & 93.76(800) & 96.50(all)  \\
DNN2(PoseSy) & -          & 75.12(300) & 81.70(600) & 85.58(900)  \\
Multiscale   & 88.62(150) & 95.14(300) & 97.69(500) & 98.51(1000)  \\
Weakly       & 91.79(171) & 93.73(343) & 93.41(686) & 94.02(915)  \\
CNN+metrix   & 82.29(200) & -          & 95.47(500) & 98.93(1000) \\
WGAN-GP      & 80.37(216) & 85.58(324) & 89.30(648) & 91.65(1080) \\
DCGAN        & -          & -          & 95.68(550) & 98.55(1099) \\
MGAN         & 85.23(200) & 90.82(400) & 94.91(800) & 97.81(all)  \\
SSDTL        & -          & -          & 90.82(550) & 93.07(1099) \\
IGAN         & 88.36(200) & 94.83(400) & 96.39(600) & -           \\
The proposed & 96.44(214) & 97.13(324) & 97.74(412) & 98.22(817) \\ \hline  \hline 
\end{tabular}}
\end{center}
\end{table}

The recognition performances are listed in Table \ref{recognition} under SOC. In Table \ref{recognition}, the number in parentheses is the number of the training samples for each method, the numbers of the labeled SAR images used for training the recognition networks are denoted as a range of numbers and the exact numbers of the labeled are marked between parentheses after the recognition rates.

From Table \ref{recognition}, it is clear that our proposed method outperforms the others with a limited sample size under SOC. In particular, under the condition that the total training samples are only 214, our method can generate effective SAR images for the recognition and still has a recognition rate of 96.44\%, which is a significant improvement compared to the recognition performance of other methods under the condition that the training samples are around 200. Therefore, it can conclude that our proposed algorithm is superior to other augmentation methods in the SAR image generation or augmentation.

\section{Conclusion}
In this paper, the proposed azimuth-controllable SAR target image generation network works for the problem of insufficient SAR target images, which is proved by the experimental results. Through the specific topological structure, the generator extracts and fuses optimally the target feature to acquire the precise generated SAR target images with the optimization information of similarity and azimuth distance provided by the similarity discriminator and azimuth predictor. The proposed azimuth-controllable SAR target image generation network obtain the capability of generating precise SAR images from two input SAR images and the azimuths of the generated SAR images can be controlled by the azimuths of the two given SAR images.

Extensive experiments have been carried out on the MSTAR dataset, and the results show clearly that not only the generated images by the proposed network are similar to the real SAR images, but also the azimuth of the generated SAR images is controllable. Besides, the generated SAR images can greatly benefit the performance of SAR ATR, especially in small sample situation. By employing SAR dataset of different imaging conditions and research demands with the proposed azimuth-controllable SAR target image generation method, it can make some contributions to the practical development of most SAR researches.

\bibliographystyle{IEEEtran}
\bibliography{ref,ref_add}

\newpage

\begin{IEEEbiography}[{\includegraphics[width=1in,height=1.25in,clip,keepaspectratio]{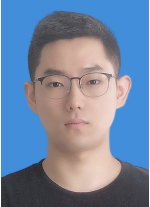}}]{Chenwei Wang}
received the B.S. degree from the School of Electronic Engineering, University of Electronic Science and Technology of China (UESTC), Chengdu, China, in 2018. He is currently pursuing the Ph.D. degree with the School of Information and Communication Engineering, University of Electronic Science and Technology of China, Chengdu, China. 
His research interests include radar signal processing, machine learning, and automatic target recognition.
\end{IEEEbiography}

\begin{IEEEbiography}[{\includegraphics[width=1in,height=1.25in,clip,keepaspectratio]{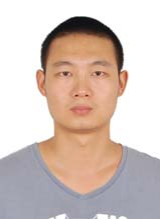}}]{Jifang Pei}
(S'12-M'19) received the B.S. degree from the College of Information Engineering, Xiangtan University, Hunan, China, in 2010, and the M.S. degree from the School of Electronic Engineering, University of Electronic Science and Technology of China (UESTC), Chengdu, China, in 2013. He received the Ph.D. degree from the School of Information and Communication Engineering, UESTC, in 2018. From 2016 to 2017, he was a joint Ph.D. Student with the Department of Electrical and Computer Engineering, National University of Singapore, Singapore. He is currently an Associate Research Fellow with the School of Information and Communication Engineering, UESTC. 
His research interests include radar signal processing, machine learning, and automatic target recognition. (Corresponding author)
\end{IEEEbiography}

\begin{IEEEbiography}[{\includegraphics[width=1in,height=1.25in,clip,keepaspectratio]{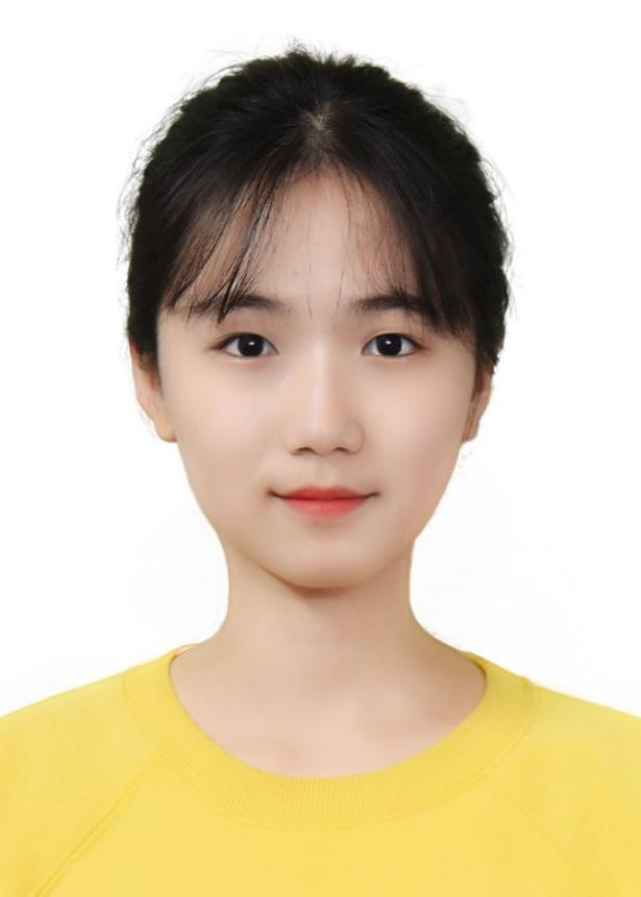}}]{Xiaoyu Liu}
received her B.S. dual degree from the University of Glasgow (UoG) and the University of Electronic Science and Technology of China (UESTC), Chengdu, China, in 2021. She is currently pursuing the M.S. degree at the School of Information and Communication Engineering, University of Electronic Science and Technology of China, Chengdu, China. 
Her research interests include machine learning, target detection, and automatic target recognition.
\end{IEEEbiography}

\begin{IEEEbiography}[{\includegraphics[width=1in,height=1.25in,clip,keepaspectratio]{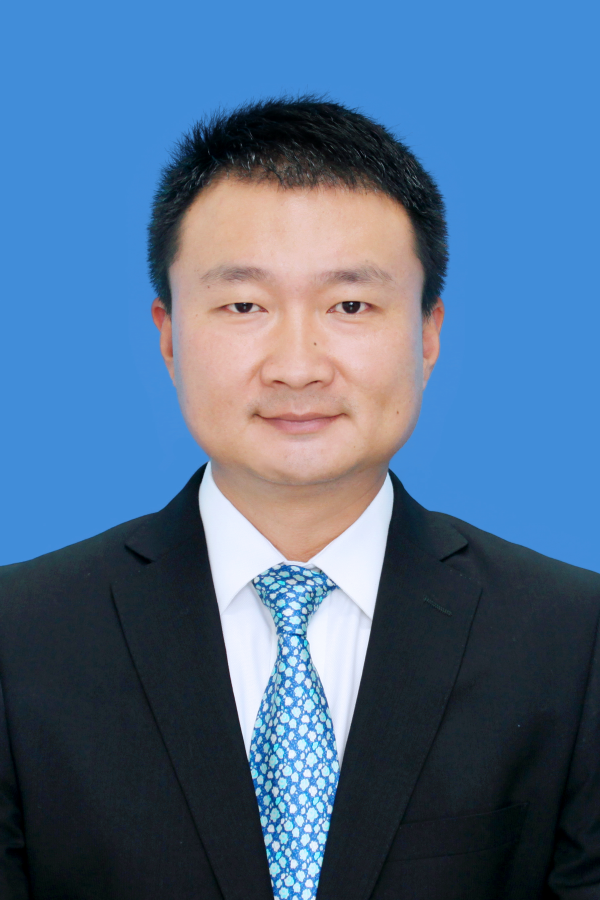}}]{Yulin Huang}
(M'08-SM'16) received the B.S. and Ph.D. degrees from the School of Electronic Engineering, University of Electronic Science and Technology of China (UESTC), Chengdu, China, in 2002 and 2008, respectively. He is currently a Professor with the UESTC. 
His research interests include radar signal processing and SAR automatic target recognition.
\end{IEEEbiography}

\begin{IEEEbiography}[{\includegraphics[width=1in,height=1.25in,clip,keepaspectratio]{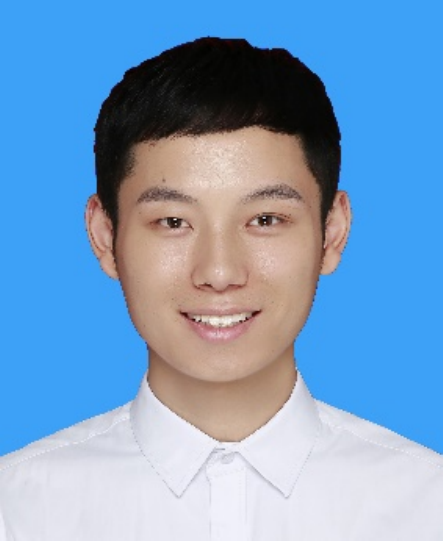}}]{Deqing Mao}
(S'17-M’22) received the B.S. degree from the School of Electronic Engineering, Chengdu University of Information Technology, Chengdu, China, in 2014. He received his Ph.D. degree with the School of Information and Communication Engineering, University of Electronic Science and Technology of China (UESTC), Chengdu, China, in 2022. From 2020 to 2021, he has been a visiting Ph.D. student at Technology University of Delft (TUD), Delft, the Netherlands. 
His research interests include radar signal processing and inverse problem in radar imaging.

\end{IEEEbiography}

\begin{IEEEbiography}[{\includegraphics[width=1in,height=1.25in,clip,keepaspectratio]{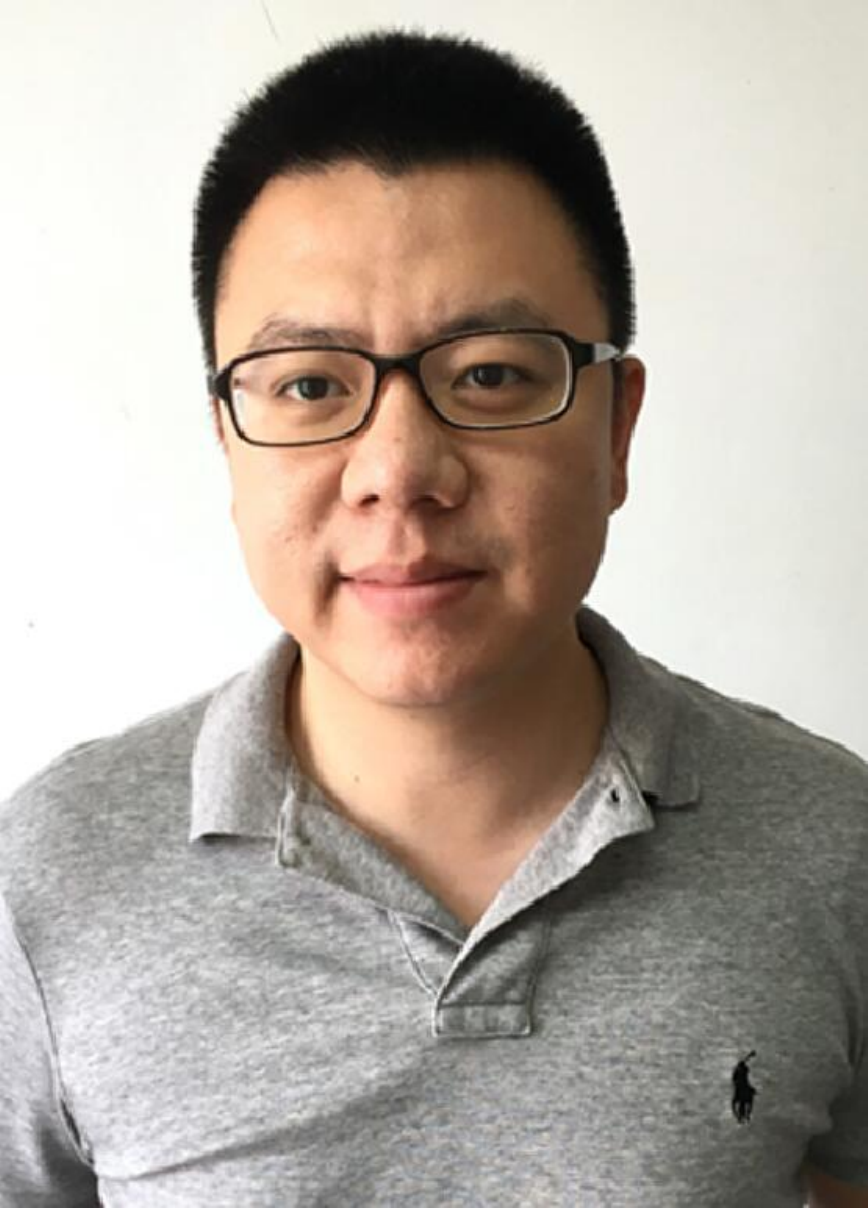}}]{Yin Zhang}
(M'16) received the B.S. and Ph.D. degrees from the School of Electronic Engineering, University of Electronic Science and Technology of China (UESTC), Chengdu, China, in 2008 and 2016, respectively. From September 2014 to September 2015, he had been a Visiting Student with the Department of Electrical and Computer Engineering, University of Delaware, Newark, DE, USA. He is currently a Research Fellow with the UESTC. 
His research interests include signal processing and radar imaging.
\end{IEEEbiography}

\begin{IEEEbiography}[{\includegraphics[width=1in,height=1.25in,clip,keepaspectratio]{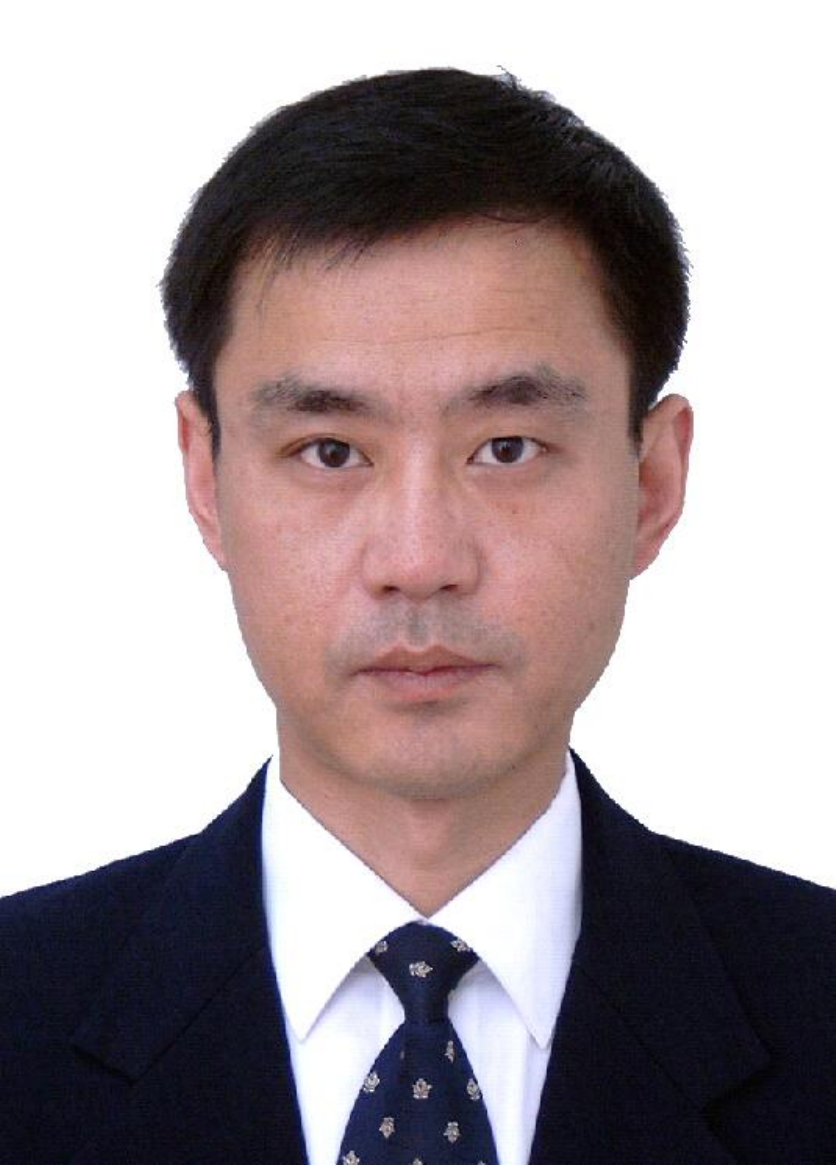}}]{Jianyu Yang}
(M'06) received the B.S. degree from the National University of Defense Technology, Changsha, China, in 1984, and the M.S. and Ph.D. degrees from the University of Electronic Science and Technology of China (UESTC), Chengdu, China, in 1987 and 1991, respectively. He is currently a Professor with the UESTC. From 2001 to 2005, he served as the Dean of School of Electronic Engineering of UESTC. In 2005, he was a Senior Visiting Scholar with the Massachusetts Institute of Technology (MIT), Cambridge, MA, USA. He was selected as the Vice-Chairman of the Radar Society of the Chinese Institute of Electronics (CIE) in 2016 and a Fellow of CIE in 2018. He serves as a Senior Editor for the Chinese Journal of Radio Science and the Journal of Systems Engineering and Electronics. 
His research interests include synthetic aperture radar imaging and automatic target recognition.
\end{IEEEbiography}

\vfill

\end{document}